\documentclass[12pt]{article}
\usepackage[T1]{fontenc}
\usepackage[latin1]{inputenc}
\usepackage{a4wide}
\usepackage{graphics}
\usepackage{psfrag}
\usepackage[dvips]{graphicx}
\usepackage{amssymb}
\usepackage{subfigure}
\makeatletter

\newcommand{\be}{\begin{equation}}
\newcommand{\ee}{\end{equation}}
\newcommand{\ba}{\begin{array}}
\newcommand{\ea}{\end{array}}
\newcommand{\bea}{\begin{eqnarray}}
\newcommand{\eea}{\end{eqnarray}}

\makeatother

\begin{document}

\makeatletter

\renewcommand\theequation{\hbox{\normalsize\arabic{section}.\arabic{equation}}} 

\@addtoreset{equation}{section} 

\renewcommand\thefigure{\hbox{\normalsize\arabic{section}.\arabic{figure}}} 

\@addtoreset{figure}{section} 

\renewcommand\thetable{\hbox{\normalsize\arabic{section}.\arabic{table}}} \@addtoreset{table}{section}

\makeatother

\title{{\normalsize \begin{flushright}\normalsize{ITP--Budapest Report
No. 573}\end{flushright}\vspace{1cm}}
Finite size effects in boundary sine-Gordon theory}

\author{Z. Bajnok, L. Palla and G. Tak{\'a}cs}

\maketitle

{\centering \emph{Institute for Theoretical Physics }\\
\emph{Roland E\"otv\"os University, }\\
\emph{H-1117 Budapest, P\'azm\'any P. s\'et\'any 1/A, Hungary}\par}

\begin{abstract}
We examine the finite volume spectrum and boundary energy in boundary
sine-Gordon theory, based on our recent results obtained by closing
the boundary bootstrap. The spectrum and the reflection
factors are checked against truncated conformal space, together with a (still
unpublished) prediction by Al.B. Zamolodchikov for the boundary energy
and the relation between the parameters of the scattering amplitudes
and of the perturbed CFT Hamiltonian. In addition, a
derivation of Zamolodchikov's formulae is given. We find an entirely consistent
picture and strong evidence for the validity of the conjectured
spectrum and scattering amplitudes, which together give a complete
description of the boundary sine-Gordon theory on mass shell.
\end{abstract}

\section{\label{sec:introduction} Introduction}

Sine-Gordon field theory is one of the most important quantum field theoretic
models with numerous applications ranging from particle theoretic problems to
condensed matter systems, and one which has played a central role in our understanding
of \( 1+1 \) dimensional field theories. A crucial property of the model is
integrability, which permits an exact analytic determination of many of its
physical properties and characteristic quantities. Integrability can
also be preserved in the presence of a boundary if suitable boundary
conditions are imposed \cite{Skl}.

In this paper, continuing our work started in \cite{neumann,bajnok1},
we investigate sine-Gordon field theory on the half-line and on a
finite volume interval, with integrable boundary conditions. It was
first pointed out by Ghoshal and Zamolodchikov \cite{ghoshal} that the
most general integrable boundary potential depends on two
parameters. They also introduced the notion of \lq boundary crossing
unitarity', and combining it with the boundary version of the Yang
Baxter equations they were able to determine soliton reflection
factors on the boundary; later Ghoshal completed this work by
determining the breather reflection factors \cite{ghoshal1} using a
boundary bootstrap equation first proposed by Fring and K{\"o}berle
\cite{FK}.

The results of Ghoshal and Zamolodchikov concerned only the reflection
factors on the ground state boundary, although they already noticed
that there are poles in the amplitudes which signal the existence of
excited boundary states. The first (partial) results on the spectrum
of these boundary states were obtained by Saleur and Skorik for
Dirichlet boundary conditions \cite{skorik}. However, they did not
take into account the boundary analogue of the Coleman-Thun mechanism,
the importance of which was first emphasized by Dorey
et. al. \cite{bct}. Using this mechanism Mattsson and Dorey were able
to close the bootstrap in the Dirichlet case and determine the
complete spectrum and the reflection factors on the excited boundary
states \cite{mattsson}. Recently we used their ideas to obtain the
spectrum of excited boundary states and their reflection factors for
the Neumann boundary condition \cite{neumann} and then for the general
two-parameter family of integrable boundary conditions
\cite{bajnok1}. For the Neumann case, extensive checks were performed
using a boundary version of the so-called Truncated Conformal Space
Approach (TCSA) \cite{YZ,dptw}; for the generic case, however, these
checks were not carried out at that time.

Another interesting problem is that of the boundary energy. Namely,
the boundary contributes a volume independent (constant) term to the
free energy, in addition to the bulk energy density which gives a term
proportional to the spatial volume. Just as in the case of the bulk
energy density, the boundary energy in general QFT is not a universal
quantity. However, in perturbed conformal field theories there is a
preferred normalization\footnote{In this preferred normalization, the
perturbing bulk and boundary operators transform homogeneously under
scale transformations.} of the Hamiltonian which gives a unique
definition for both the bulk and the boundary
contributions. Therefore, this boundary energy is an interesting
quantity to compute. For Dirichlet boundary condition it was obtained
by Leclair et al. in \cite{leclair}. A few years ago
Al. B. Zamolodchikov presented a result for general integrable
boundary conditions \cite{unpublished}.

One crucial ingredient, that is needed e.g. for a TCSA check of the
spectrum and reflection factors for the general integrable boundary
conditions, is a relation between the ultraviolet (UV) parameters that
appear in the perturbed CFT Hamiltonian and the infrared (IR)
parameters in the reflection factors. This relation was also obtained
by Al. B. Zamolodchikov \cite{unpublished}. Using his result, we
perform an extensive check of the spectrum, boundary energy and
reflection factors of boundary sine-Gordon theory. This provides
strong evidence that all the results mentioned above form a
consistent and complete description of the boundary sine-Gordon theory
on mass shell (i.e. spectrum and scattering amplitudes).

The paper is organized as follows. In Section
\ref{sec:boundary_bootstrap} we recall the results on the boundary
bootstrap in boundary sine-Gordon theory.  Section \ref{sec: UV-IR}
describes Zamolodchikov's formulae on the UV-IR relation and the
boundary energy. These formulae were presented at some seminars, but
have not been published; however, we could get some notes taken by the
audience.\footnote{We thank P. Dorey and G. Watts for making the notes
available to us.}  In these notes we found several misprints; in order
to determine the correct form of the formulae (which is of utmost
importance in order to make the comparison to TCSA), we rederive here
the boundary energy using the thermodynamic Bethe Ansatz (TBA) and
then check the UV-IR relation using the exact vacuum expectation
values of boundary fields conjectured by Fateev, Zamolodchikov and
Zamolodchikov \cite{FZZ}.\footnote{The derivation presented here
is very similar to the way Al.B. Zamolodchikov arrived to the formulae
(\ref{UV_IR}-\ref{boundary_energy}) in Section 3.1, according to the
hints he gave in his seminars.} In Section \ref{sec:general_TCSA} we
describe the results coming from TCSA for generic (non Dirichlet)
boundary conditions, while in Section \ref{sec:dirichlet_TCSA} we
present the results for Dirichlet boundary conditions, which are a
singular limit of the generic case and so the TCSA must be set up
differently. The paper ends with some brief conclusions and an outlook in
Section \ref{sec:conclusions}.

\section{\label{sec:boundary_bootstrap} Boundary bootstrap in sine-Gordon theory }

Boundary sine-Gordon theory is defined by the action
\begin{equation}
\label{bsg_action}
\mathcal{A}_{sG}=\int ^{\infty }_{-\infty }dt\left( \int ^{0}_{-\infty }dx\left[ \frac{1}{2}\partial _{\mu }\Phi \partial ^{\mu }\Phi +\frac{m^{2}_{0}}{\beta ^{2}}\cos \beta \Phi \right] +M_{0}\cos \frac{\beta }{2}\left( \Phi (0,\, t)-\phi _{0}\right) \right) 
\end{equation}
where \( \Phi (x,\, t) \) is a real scalar field and \( M_0\),
\(\phi_0\) are the two parameters characterizing the boundary
condition:
\begin{equation}
\label{hatfel}
\partial_x\Phi(x,t)\vert_{x=0}=-M_0\frac{\beta}{2}\sin\left(\frac{\beta}{2}
(\Phi(0,t)-\phi_0)\right).
\end{equation}
Ghoshal and Zamolodchikov showed that the above model is integrable
\cite{ghoshal} and that the above boundary potential is the most
general that permits the existence of higher spin conserved charges. 

\subsection{Bulk scattering properties}

In the bulk sine-Gordon model the particle spectrum consists of the soliton
\( s \), the antisoliton \( \bar{s} \) , and the breathers \( B^{n} \) which
appear as bound states in the \( s\bar{s} \) scattering amplitude \( S^{-+}_{+-} \)
. As a consequence of the integrable nature of the model any scattering amplitude
factorizes into a product of two particle scattering amplitudes, from which
the independent ones in the purely solitonic sector are \cite{ZZ} 
\begin{eqnarray}
a(u)= & S^{++}_{++}(u)=S_{--}^{--}(u)= & 
-\prod ^{\infty }_{l=1}\left[ \frac{\Gamma (2(l-1)\lambda-\frac{\lambda u}{\pi })
\Gamma (2l\lambda +1-\frac{\lambda u}{\pi })}
{\Gamma ((2l-1)\lambda -\frac{\lambda u}{\pi })
\Gamma ((2l-1)\lambda +1-\frac{\lambda u}{\pi })}/(u\to -u)\right] 
\nonumber \\
b(u)= & S^{+-}_{+-}(u)=S_{-+}^{-+}(u)= & 
\frac{\sin (\lambda u)}{\sin (\lambda (\pi -u))}a(u)\quad \qquad ;\qquad\nonumber \\
c(u)= & S^{-+}_{+-}(u)=S_{-+}^{+-}(u)= & 
\frac{\sin (\lambda \pi )}{\sin (\lambda (\pi -u))}a(u)\quad \qquad ;
\qquad  
\, \, \, \label{abc} 
\end{eqnarray}
where the parameter \(\lambda \) is determined by the sine-Gordon
coupling constant 
\be\label{lalam} \lambda =\frac{8\pi }{\beta ^{2}}-1
\ee 
and
\(u=-i\theta\) denotes the purely imaginary rapidity. 
The other scattering amplitudes can be described in terms of the
functions
\[
\{y\}=\frac{\left( \frac{y+1}{2\lambda }\right) \left(
\frac{y-1}{2\lambda }\right) }{\left( \frac{y+1}{2\lambda }-1\right)
\left( \frac{y-1}{2\lambda }+1\right) }\quad ,\quad (x)=\frac{\sin
\left( \frac{u}{2}+\frac{x\pi }{2}\right) }{\sin \left(
\frac{u}{2}-\frac{x\pi }{2}\right) }\, \, \, ,\, \, \,
\{y\}\{-y\}=1\quad ,\, \, \{y+2\lambda \}=\{-y\}
\]
as follows. For the scattering of the breathers \( B^{n} \) and \( B^{m} \)
with \( n\geq m \) and relative rapidity \( u \) the amplitude takes the form
\[
S^{n\, m}(u)=S^{n\, m}_{n\, m}(u)=\{n+m-1\}\{n+m-3\}\dots
\{n-m+3\}\{n-m+1\}\, \, \, ,
\]
while for the scattering of the soliton (antisoliton) and \( B^{n} \) 
\[
S^{n}(u)=S_{+\, n}^{+\, n}(u)=S_{-\, n}^{-\, n}(u)=\{n-1+\lambda \}\{n-3+\lambda \}\dots \left\{ \begin{array}{c}
\{1+\lambda \}\quad \textrm{if }n\textrm{ is even}\\
-\sqrt{\{\lambda \}}\quad \textrm{if }n\textrm{ is odd}\, \, \, .
\end{array}\right. 
\]

\subsection{Ground state reflection factors}

The most general reflection factor - modulo CDD-type factors - of the soliton
antisoliton multiplet \( |s,\bar{s}\rangle  \) on the ground state boundary,
denoted by \( |\, \rangle  \), satisfying the boundary versions of the Yang
Baxter, unitarity and crossing equations was found by Ghoshal and Zamolodchikov
\cite{ghoshal}:\begin{eqnarray}
R(\eta ,\vartheta ,u) & = & \left( \begin{array}{cc}
P^{+}(\eta ,\vartheta ,u) & Q(\eta ,\vartheta ,u)\\
Q(\eta ,\vartheta ,u) & P^{-}(\eta ,\vartheta ,u)
\end{array}\right) \nonumber \\
 & = & \left( \begin{array}{cc}
P_{0}^{+}(\eta ,\vartheta ,u) & Q_{0}(u)\\
Q_{0}(u) & P_{0}^{-}(\eta ,\vartheta ,u)
\end{array}\right) R_{0}(u)\frac{\sigma (\eta ,u)}{\cos (\eta )}\frac{\sigma (i\vartheta ,u)}{\cosh (\vartheta )}\, \, \, ,\nonumber \\
P_{0}^{\pm }(\eta ,\vartheta ,u) & = & \cos (\lambda u)\cos (\eta )\cosh (\vartheta )\mp \sin (\lambda u)\sin (\eta )\sinh (\vartheta )\nonumber \\
Q_{0}(u) & = & -\sin (\lambda u)\cos (\lambda u)\label{Rsas} 
\end{eqnarray}
 where \( \eta  \) and \( \vartheta  \) are the two real parameters characterizing
the solution, \[
R_{0}(u)=\prod ^{\infty }_{l=1}\left[ \frac{\Gamma (4l\lambda -\frac{2\lambda u}{\pi })\Gamma (4\lambda (l-1)+1-\frac{2\lambda u}{\pi })}{\Gamma ((4l-3)\lambda -\frac{2\lambda u}{\pi })\Gamma ((4l-1)\lambda +1-\frac{2\lambda u}{\pi })}/(u\to -u)\right] \]
 is the boundary condition independent part and \[
\sigma (x,u)=\frac{\cos x}{\cos (x+\lambda u)}\prod ^{\infty }_{l=1}\left[ \frac{\Gamma (\frac{1}{2}+\frac{x}{\pi }+(2l-1)\lambda -\frac{\lambda u}{\pi })\Gamma (\frac{1}{2}-\frac{x}{\pi }+(2l-1)\lambda -\frac{\lambda u}{\pi })}{\Gamma (\frac{1}{2}-\frac{x}{\pi }+(2l-2)\lambda -\frac{\lambda u}{\pi })\Gamma (\frac{1}{2}+\frac{x}{\pi }+2l\lambda -\frac{\lambda u}{\pi })}/(u\to -u)\right] \]
describes the boundary condition dependence. Note that the topological charge
may be changed by two in these reflections, thus the parity of the soliton number
is conserved. 

As a consequence of the bootstrap equations \cite{ghoshal} the breather reflection
factors share the structure of the solitonic ones, \cite{ghoshal1}: \begin{equation}
\label{Rbr1}
R^{(n)}(\eta ,\vartheta ,u)=R_{0}^{(n)}(u)S^{(n)}(\eta ,u)S^{(n)}(i\vartheta ,u)\, \, \, ,
\end{equation}
where \begin{equation}
\label{Rbr2}
R_{0}^{(n)}(u)=\frac{\left( \frac{1}{2}\right) \left( \frac{n}{2\lambda }+1\right) }{\left( \frac{n}{2\lambda }+\frac{3}{2}\right) }\prod ^{n-1}_{l=1}\frac{\left( \frac{l}{2\lambda }\right) \left( \frac{l}{2\lambda }+1\right) }{\left( \frac{l}{2\lambda }+\frac{3}{2}\right) ^{2}}\quad ;\quad S^{(n)}(x,u)=\prod ^{n-1}_{l=0}\frac{\left( \frac{x}{\lambda \pi }-\frac{1}{2}+\frac{n-2l-1}{2\lambda }\right) }{\left( \frac{x}{\lambda \pi }+\frac{1}{2}+\frac{n-2l-1}{2\lambda }\right) }\, \, \, .
\end{equation}
 In general \( R_{0}^{(n)} \) describes the boundary independent properties
and the other factors give the boundary dependent ones.

\subsection{The spectrum of boundary bound states and the associated reflection factors}

In the general case, the spectrum of boundary bound states was derived in \cite{bajnok1}.
It is a straightforward generalization of the spectrum in the Dirichlet limit
previously obtained by Mattsson and Dorey \cite{mattsson}. The states can be
labeled by a sequence of integers \( |n_{1},n_{2},\dots ,n_{k}\rangle  \).
Such a state exists whenever the \[
\frac{\pi }{2}\geq \nu _{n_{1}}>w_{n_{2}}>\nu _{n_{3}}>w_{n_{4}}>\dots \geq 0\]
 condition holds, where \[
\nu _{n}=\frac{\eta }{\lambda }-\frac{(2n+1)\pi }{2\lambda }\; 
\qquad\mathrm{and}\;\qquad  w_{n}=\pi -\frac{\eta }{\lambda }-\frac{(2n-1)\pi }{2\lambda }\; \]
denote the location of certain poles in $\sigma (\eta ,u)$. 
 The mass of such a state (i.e. its energy above the ground state) is 

\begin{equation}
m_{|n_{1},n_{2},\dots ,n_{k}\rangle }=M\sum _{i\textrm{
}\mathrm{odd}}\cos (\nu _{n_{i}})+M\sum _{i\textrm{
}\mathrm{even}}\cos (w_{n_{i}})\, \, \, ,\label{excited_energy}
\end{equation}
where $M$ is the soliton mass.
The reflection factors of the various particles on these boundary
states 
depend on whether $k$ is even or odd. When \( k \) is even, the reflection factors take the form\[
Q_{|n_{1},n_{2},\dots ,n_{k}\rangle }(\eta ,\vartheta ,u)=Q(\eta ,\vartheta ,u)\prod _{i\textrm{ }\mathrm{odd}}a_{n_{i}}(\eta ,u)\prod _{i\textrm{ }\mathrm{even}}a_{n_{i}}(\bar{\eta },u)\, \, \, ,\]
and \[
P^{\pm }_{|n_{1},n_{2},\dots ,n_{k}\rangle }(\eta ,\vartheta ,u)=P^{\pm }(\eta ,\vartheta ,u)\prod _{i\textrm{ }\mathrm{odd}}a_{n_{i}}(\eta ,u)\prod _{i\textrm{ }\mathrm{even}}a_{n_{i}}(\bar{\eta },u)\, \, \, ,\]
for the solitonic processes, where\[
a_{n}(\eta ,u)=\prod _{l=1}^{n}\left\{ 2\left( \frac{\eta }{\pi }-l\right) \right\} \quad ;\quad \bar{\eta }=\pi (\lambda +1)-\eta \, \, \, .\]
 For the breather reflection factors the analogous formula is 
\be\label{Rbe1}
R^{(n)}_{|n_{1},n_{2},\dots ,n_{k}\rangle }(\eta ,\vartheta ,u)=R^{(n)}(\eta ,\vartheta ,u)\prod _{i\textrm{ }\mathrm{odd}}b^{n}_{n_{i}}(\eta ,u)\prod _{i\textrm{ }\mathrm{even}}b^{n}_{n_{i}}(\bar{\eta },u)\, \, \, \ee
where now \be\label{Rbe2}
b_{k}^{n}(\eta ,u)=\prod _{l=1}^{\min (n,k)}\left\{ \frac{2\eta }{\pi }-\lambda +n-2l\right\} \left\{ \frac{2\eta }{\pi }+\lambda -n-2(k+1-l)\right\} \, \, \, .\ee
In the case when \( k \) is odd, the same formulae apply if in
the \(P^\pm\), \(Q\) and \(R^{(n)}\) 
ground state reflection factors the \( \eta \leftrightarrow \bar{\eta } \)
and \( s\leftrightarrow \bar{s} \) changes are made.

\section{\label{sec: UV-IR} Boundary energy and UV-IR relation in sine-Gordon theory}

\subsection{Zamolodchikov's formulae}

As mentioned in the introduction, recently Al. B. Zamolodchikov
presented (but not yet published) \cite{unpublished} a formula for the
relation between the UV and the IR parameters in the sine-Gordon
model. To set the conventions for this relation, consider
boundary sine-Gordon theory as a joint bulk and boundary perturbation
of the \( c=1 \) free boson with Neumann boundary conditions
(perturbed conformal field theory, pCFT):\begin{equation}
\label{pCFT_action}
\mathcal{A}_{pCFT}=\mathcal{A}^{N}_{c=1}+\mu \, \int ^{\infty }_{-\infty }dt\int ^{0}_{-\infty }dx\, :\cos \beta \Phi (x,\, t):+\tilde{\mu }\, \int ^{\infty }_{-\infty }dt\, :\cos \frac{\beta }{2}\left( \Phi (0,\, t)-\phi _{0}\right) :
\end{equation}
 where the colons denote the standard CFT normal ordering, which defines the
normalization of the operators and of the coupling constants. The couplings
\( \mu  \) and \( \tilde{\mu } \) have nontrivial dimensions; 
\[
[\mu ]=[ {\rm mass} ]^{2-2h_\beta},\qquad [\tilde{\mu }]=[ {\rm mass}
]^{1-h_\beta},\qquad h_\beta =\frac{\beta^2}{8\pi},
\] 
see the section
on TCSA for more details. With these conventions the UV-IR relation
\footnote{A similar relation was derived by Corrigan and Taormina
\cite{Ed} for sinh-Gordon theory, however, their normalization of the
coupling constants is different from the one natural in the perturbed
CFT framework.
} is
\begin{eqnarray}\label{UV_IR}
\cos \left( \frac{\beta ^{2}\eta }{8\pi }\right) \cosh \left( \frac{\beta ^{2}\vartheta }{8\pi }\right)  & = & \frac{\tilde{\mu }}{\tilde{\mu }_{\mathrm{crit}}}\cos \left( \frac{\beta \phi _{0}}{2}\right)\,, \nonumber \\
\sin \left( \frac{\beta ^{2}\eta }{8\pi }\right) \sinh \left( \frac{\beta ^{2}\vartheta }{8\pi }\right)  & = & \frac{\tilde{\mu }}{\tilde{\mu }_{\mathrm{crit}}}\sin \left( \frac{\beta \phi _{0}}{2}\right)\,, \label{UV_IR_relation} 
\end{eqnarray}
where \begin{equation}
\label{mu_crit}
\tilde{\mu }_{\mathrm{crit}}=\sqrt{\frac{2\mu }{\sin \left(
\frac{\beta ^{2}}{8}\right) }}\ \ .
\end{equation}
Zamolodchikov also gave the boundary energy as
\begin{equation}
\label{boundary_energy}
E(\eta ,\vartheta )=-\frac{M}{2\cos \frac{\pi }{2\lambda }}\left( \cos \left( \frac{\eta }{\lambda }\right) +\cosh \left( \frac{\vartheta }{\lambda }\right) -\frac{1}{2}\cos \left( \frac{\pi }{2\lambda }\right) +\frac{1}{2}\sin \left( \frac{\pi }{2\lambda }\right) -\frac{1}{2}\right)\,. 
\end{equation}

\subsection{Derivation of the boundary energy from TBA}

In an integrable boundary theory with one scalar particle of mass \( m \) only,
one can write down the TBA equation for the ground state energy on a strip with
spatial volume \( L \) and integrable boundary conditions \( a \) and \( b \)
at the two ends. The equation is of the form \cite{leclair}:
\begin{equation}
\label{generic_TBA}
\varepsilon \left( \theta \right) =2l\cosh \theta +k_{ab}\left( \theta \right) -\int ^{\infty }_{-\infty }\frac{d\theta '}{2\pi }\varphi \left( \theta -\theta '\right) \log \left( 1+\mathrm{e}^{-\varepsilon \left( \theta '\right) }\right) 
\end{equation}
where \( l=mL \) is the dimensionless volume parameter. The kernel is
expressed in terms of the two-body \( S \)-matrix \( S(\theta ) \) as  
\[
\varphi (\theta )=-i\frac{\partial }{\partial \theta }\log S\left( \theta \right)\,, \]
while\[
k_{ab}(\theta )=-\log \left[ R_{a}\left( \frac{i\pi }{2}-\theta \right) R_{b}\left( \frac{i\pi }{2}+\theta \right) \right]\,, \]
where \( R_{a}\left( \theta \right)  \) and \( R_{b}\left( \theta \right)  \)
are the reflection factors for the two ends. From the solution \( \varepsilon \left( \theta \right)  \)
of the TBA equation the ground state energy can be calculated using the formula
\begin{equation}
\label{tba_energy}
E(L)=E_{\mathrm{bulk}}L+E_{\mathrm{boundary}}-\frac{\pi c(mL)}{24L}\;
,\quad c(l)=\frac{6l}{\pi ^{2}}\int ^{\infty }_{-\infty }d\theta \:
L(\theta )\cosh \theta \ ,
\end{equation}
where \(L(\theta )\) is the usual short hand notation \( L(\theta
)=\log \left( 1+\mathrm{e}^{-\varepsilon \left( \theta \right)
}\right)\).  It is well-known that no such TBA equation (or, for that
matter, a finite system of TBA equations) can be written for
sine-Gordon theory as a result of the nondiagonal bulk and boundary
scattering of the solitons (except for special values of the
parameters). Therefore, our approach is to calculate the boundary
energy for sinh-Gordon theory and then analytically continue back to
the sine-Gordon case. This is known to work e.g. for \( S \)-matrices,
form factors and many other quantities, and so we simply assume it
works for the boundary energy as well.

Consider therefore the boundary energy in boundary sinh-Gordon theory\[
\mathcal{A}_{shG}=\int ^{\infty }_{-\infty }dt\left( \int ^{0}_{\infty }dx\left[ \frac{1}{2}\partial _{\mu }\Phi \partial ^{\mu }\Phi -\frac{m^{2}_{0}}{b^{2}}\cosh b\Phi \right] -M_{0}\cosh \frac{b}{2}\left( \Phi -\phi_{0}\right) \right)\,, \]
which can be considered as the analytic continuation of the boundary sine-Gordon
model (\ref{bsg_action}) by substituting \( b=i\beta  \) (and changing the
convention for the sign of \( M_{0} \)). Then from (\ref{lalam})\[
\lambda =-\frac{8\pi }{b^{2}}-1\]
and, as a result, \( \lambda  \) is negative for the sinh-Gordon case. Note
that the analytic continuation is through the point \( \lambda =\infty  \)
(complex infinity), therefore for the purposes of relating physical quantities
between the two models the natural variable is \( \lambda ^{-1} \).

We now proceed to the calculation of the boundary energy. A similar
calculation was performed by Dorey at al. \cite{dptw} for the scaling
Lee-Yang case. They presented the general idea with enough hints to
reconstruct the method, but for the sake of completeness we write
down the details for the interested reader. It is based on Zamolodchikov's
method for obtaining the bulk energy from the TBA with periodic
boundary conditions \cite{yl_potts_tba}.

Suppose for simplicity that the boundary conditions \( a \) and \( b \)
are identical and so \( k=k_{aa} \) is even. Then in general the functions \( k \)
and \( \varphi  \) have the following asymptotic behaviour\begin{eqnarray}
k(\theta ) & \sim  & k_{0}+A\mathrm{e}^{-\left| \theta \right| }+\dots \nonumber \\
\varphi \left( \theta \right)  & \sim  & C\mathrm{e}^{-\left| \theta \right| }+\dots \label{scatt_asymp} 
\end{eqnarray}
for \( \left| \theta \right| \, \rightarrow \, \infty  \), where \( k_{0} \),
\( A \) and \( C \) are real constants. 

The `kink' functions, defined as
\[
\varepsilon _{\pm }\left( \theta \right) =\lim _{l\, \rightarrow \, 0}\varepsilon \left( \theta \pm \log \frac{1}{l}\right) \,,
\]
satisfy the `kink' equation\[
\varepsilon _{\pm }\left( \theta \right) =\mathrm{e}^{\pm \theta }+k_{0}-\int ^{\infty }_{-\infty }\frac{d\theta '}{2\pi }\varphi \left( \theta -\theta '\right) \log \left( 1+\mathrm{e}^{-\varepsilon _{\pm }\left( \theta '\right) }\right) \]
and are related as \[
\varepsilon _{-}\left( \theta \right) =\varepsilon _{+}\left( -\theta \right)\,. \]
Let us also introduce the following definitions\[
 L_{\pm }\left( \theta \right) =
\log \left( 1+\mathrm{e}^{-\varepsilon _{\pm }\left( \theta \right) }\right) \]
and define the asymptotic values\[
\varepsilon _{0}=\varepsilon _{+}(-\infty )\quad ,\quad L_{0}=L_{+}(-\infty )=\log \left( 1+\mathrm{e}^{-\varepsilon _{0}}\right) \]
which satisfy the standard `plateau' equation\begin{equation}
\label{plateau_eqn}
\varepsilon _{0}=k_{0}-NL_{0}\quad ,\quad N=\int ^{\infty }_{-\infty
}\frac{d\theta }{2\pi }\varphi \left( \theta \right)\ . 
\end{equation}
Our aim is to expand \( c(l) \) around \( l=0 \). To calculate the
first few terms, it is convenient to define the functions \( \delta \) and \( \tilde{L}
\) in the following way:
\begin{eqnarray}
\varepsilon (\theta ) & = & \varepsilon _{+}\left( \theta -\log \frac{1}{l}\right) +\varepsilon _{+}\left( -\theta -\log \frac{1}{l}\right) +\delta \left( \theta \right) -\varepsilon _{0}\,,\nonumber \\
L(\theta ) & = & L_{+}\left( \theta -\log \frac{1}{l}\right) +L_{+}\left( -\theta -\log \frac{1}{l}\right) +\tilde{L}\left( \theta \right) -L_{0}\,.\label{kink_splitting} 
\end{eqnarray}
They satisfy\begin{eqnarray}
\delta \left( \theta \right)  & = & k\left( \theta \right) -k_{0}-\int ^{\infty }_{-\infty }\frac{d\theta '}{2\pi }\varphi \left( \theta -\theta '\right) \tilde{L}\left( \theta '\right)\,, \label{l_eta_eqn} \\
 &  & \delta \left( \theta \right) \, ,\, \tilde{L}\left( \theta \right) \, \rightarrow \, 0\; \mathrm{as}\; l\, \rightarrow \, 0\nonumber .
\end{eqnarray}
We can then rewrite
\[
c(l)=\frac{6}{\pi ^{2}}\int ^{\infty }_{-\infty }d\theta \: \mathrm{e}^{\theta }L_{+}(\theta )+\frac{6l^{2}}{\pi ^{2}}\int ^{\infty }_{-\infty }d\theta \: \mathrm{e}^{-\theta }\frac{\partial L_{+}}{\partial \theta }+\frac{6l}{\pi ^{2}}\int ^{\infty }_{-\infty }d\theta \, \tilde{L}\left( \theta \right) \cosh \theta \]
The first term gives the UV central charge and can be calculated using the standard
dilogarithm sum rules. The second term is the (anti) bulk energy density, that
can be calculated self-consistently by examining the \( \theta \rightarrow -\infty  \)
asymptotics of the integrand \cite{yl_potts_tba}: \[
\frac{\partial L_{+}}{\partial \theta }=-\frac{1}{1+\mathrm{e}^{\varepsilon _{+}(\theta )}}\frac{\partial \varepsilon _{+}}{\partial \theta }\sim -\frac{1}{1+\mathrm{e}^{\varepsilon _{0}}}\frac{\partial \varepsilon _{+}}{\partial \theta }\quad \mathrm{for}\quad \theta \rightarrow -\infty \]
The terms proportional to \( \mathrm{e}^{\theta } \) must cancel for the integral
to converge on its lower bound. Using the kink equation and the asymptotics
of \( \varphi  \), to leading order
\[
\frac{\partial \varepsilon _{+}}{\partial \theta }=\mathrm{e}^{\theta
}\left( 1-\frac{C}{2\pi }\int ^{\infty }_{-\infty }d\theta '\,
\mathrm{e}^{-\theta '}\frac{\partial L_{+}}{\partial \theta '}\right) 
\]
from which 
\[
\int ^{\infty }_{-\infty }d\theta \, \mathrm{e}^{-\theta }\frac{\partial L_{+}}{\partial \theta }=\frac{2\pi }{C}\,.\]
 In the perturbed conformal field theory formalism, the ground state energy
can be expanded as\[
E(L)=\frac{\pi }{L}\sum ^{\infty }_{n=0}\mathcal{C}_{n}\, (mL)^{n(1-\Delta )}\]
so the terms linear in \( L \) must cancel from (\ref{tba_energy}). Therefore
we obtain the bulk energy density as\[
E_{\mathrm{bulk}}=\frac{1}{2C}m^{2}\, .\]
The third term can be rewritten using that \( \tilde{L}\left( \theta \right) =\tilde{L}\left( -\theta \right)  \):\[
\int ^{\infty }_{-\infty }d\theta \, \tilde{L}\left( \theta \right) \cosh \theta =\int ^{\infty }_{-\infty }d\theta \, \tilde{L}\left( \theta \right) \, \mathrm{e}^{-\theta }\]
After a partially integration, it can be seen that once again, the integral is convergent
if terms proportional to \( \mathrm{e}^{\theta } \) cancel in \( \frac{\partial \tilde{L}}{\partial \theta } \).
Using equations (\ref{kink_splitting}) this is equivalent to cancellation of
all terms proportional \( \mathrm{e}^{\theta } \) in \( \delta  \), at least
to leading order in \( l \). From (\ref{l_eta_eqn}) we obtain
\[
\delta \left( \theta \right) =\mathrm{e}^{\theta }\left(
A-\frac{C}{2\pi }\int ^{\infty }_{-\infty }d\theta '\tilde{L}\left(
\theta '\right) \, \mathrm{e}^{-\theta '}\right) \] from which we
obtain (to leading order) \[ \int ^{\infty }_{-\infty }d\theta \,
\tilde{L}\left( \theta \right) \, \mathrm{e}^{-\theta }=\frac{2\pi
A}{C}\,.
\] 
None of the subleading terms contains any contribution
which are independent of the volume and therefore in
(\ref{tba_energy}) \( E_{\mathrm{boundary}} \) must cancel against
this particular term, leading to \[
E_{\mathrm{boundary}}=\frac{A}{2C}m\ . \]

\subsection{The sinh-Gordon case}

In sinh-Gordon theory the two particle \( S \)-matrix can be written
as (remember, that in our convention \(\lambda \) is negative in its
physical range):
\begin{equation}
\label{s_matrix}
S\left( \theta \right) =\frac{\sinh \theta +i\sin \frac{\pi }{\lambda
}}{\sinh \theta -i\sin \frac{\pi }{\lambda }}\ .
\end{equation}
As a result, the TBA kernel is
\[
\varphi (\theta )=-\frac{2\cosh \theta \sin \frac{\pi }{\lambda }}{\sinh ^{2}\theta +\sin ^{2}\frac{\pi }{\lambda }}\, \sim \, -4\sin \frac{\pi }{\lambda }\, \mathrm{e}^{-\left| \theta \right| }+O\left( \mathrm{e}^{-2\left| \theta \right| }\right)\,, \]
and so we get
\[
C=-4\sin \frac{\pi }{\lambda }\ .\]
The integral \( N \) takes the value\[
N=\int ^{\infty }_{-\infty }\frac{d\theta }{2\pi }\varphi \left( \theta \right) =\left\{ \begin{array}{rcl}
1 & \mathrm{for} & \Re \mathrm{e}\, \lambda <0\\
-1 & \mathrm{for} & \Re \mathrm{e}\, \lambda >0
\end{array}\right. \]
which means that the plateau equation (\ref{plateau_eqn}) has the
solution\[ \mathrm{e}^{-\varepsilon
_{0}}=\frac{\mathrm{e}^{-k_{0}}}{1+\mathrm{e}^{-k_{0}}\,
\mathrm{sign}\, \Re \mathrm{e}\, \lambda }\ .\] Note that there is no
real solution for \( \lambda <0 \), \( k_{0}\leq 0 \).  This
peculiarity of the sinh-Gordon TBA equation was already noted by Al.B.
Zamolodchikov in the case of periodic boundary condition
\cite{sinhG_tba}.  We simply assume that we are working for parameter
values for which such a solution exists, so the considerations of the
previous subsection apply. This is always the case for \( \Re
\mathrm{e}\, \lambda >0 \), which is however not a physical range of
the parameter \( \lambda \) in sinh-Gordon theory.  Therefore we treat
the sinh-Gordon TBA in this range as a mathematical problem only,
without a corresponding physical field theory (except for the case \(
\lambda =3/2 \), see later). We further assume that all physical
quantities that we wish to calculate are meromorphic functions of \(
\lambda ^{-1} \) and so they have a unique analytic continuation to
the values of \( \lambda ^{-1} \) that we are interested
in\footnote{It is clear that the relevant variable to consider is \(
\lambda ^{-1} \) because the continuation in the coupling goes through
the value \( \beta=0 \) which corresponds to \( \lambda = \infty
\)}. Note that this argument is not a proper derivation; however, for
the time being this is the only way we can arrive at the desired
result, and we show that the results fit with the bootstrap spectrum,
TCSA data and known results from previous literature.

E.g. for the bulk energy density we obtain\[
E^{\mathrm{shG}}_{\mathrm{bulk}}=-\frac{m^{2}}{8\sin \frac{\pi
}{\lambda }}\ .\]
This is meromorphic in \( \lambda ^{-1} \) and so we trust that it is the true
bulk energy constant of the sinh-Gordon theory in the regime \( \lambda <0 \).
Furthermore, it is equal to the known result \cite{sinhG_bulk}. Now we can
try and continue this result to the sine-Gordon regime \( \lambda >0 \). Under
this continuation the sinh-Gordon particle is identified with the first breather
of sine-Gordon theory and so we have\[
m=2M\sin \frac{\pi }{2\lambda }\,,\]
where \( M \) is the soliton mass. We then obtain\[
E^{\mathrm{sG}}_{\mathrm{bulk}}=-\frac{M^{2}}{4}\tan \frac{\pi }{2\lambda }\,,\]
which is the correct bulk energy density of sine-Gordon theory \cite{massgap}. 
Hence the above method of continuation works for the bulk
energy constant. 

Now let us calculate the boundary energy. From
eqns. (\ref{Rbr1},\ref{Rbr2}), the reflection factor of the first
breather in sine-Gordon theory can be written as
\begin{equation}
\label{b1_refl}
R^{(1)}(\theta )=\frac{\left( \frac{1}{2}\right) _{\theta }\left(
\frac{1}{2\lambda }+1\right) _{\theta }}{\left( \frac{1}{2\lambda
}+\frac{3}{2}\right) _{\theta }}\frac{\left( \frac{\eta }{\pi \lambda
}-\frac{1}{2}\right) _{\theta }\left( \frac{i\vartheta }{\pi \lambda
}-\frac{1}{2}\right) _{\theta }}{\left( \frac{\eta }{\pi \lambda
}+\frac{1}{2}\right) _{\theta }\left( \frac{i\vartheta }{\pi \lambda
}+\frac{1}{2}\right) _{\theta }}\quad ,\quad 
(x)_{\theta }\equiv (x)=\frac{\sinh \left( \frac{\theta
}{2}+i\frac{\pi x}{2}\right) }{\sinh \left( \frac{\theta
}{2}-i\frac{\pi x}{2}\right) }\,,
\end{equation}
where \( \eta  \) and \( \vartheta  \) parametrize the boundary conditions.
The sinh-Gordon reflection factor can be obtained by continuing the reflection
factor to negative values of \( \lambda^{-1} \) (for sinh-Gordon theory, \( \eta  \)
is real and \( \vartheta  \) is purely imaginary, while for sine-Gordon theory
both parameters are real). Putting the same boundary condition on the two boundaries
of the strip (with the same values of \( \vartheta  \) and \( \eta  \)) we
obtain\[
E^{\mathrm{shG}}_{\mathrm{boundary}}=2E^{\mathrm{shG}}(\eta ,\, \vartheta )\]
where \( E(\eta ,\, \vartheta ) \) is the energy of a single boundary. The
term \( k\left( \theta \right)  \) in the TBA equation (\ref{generic_TBA})
is \[
k\left( \theta \right) =-\log \left[K\left( \theta \right) K\left( -\theta
\right)\right] \quad ,\quad K\left( \theta \right) =R^{(1)}\left( i\frac{\pi
}{2}-\theta \right)\ . \]
Using the identity\[
(x)_{i\frac{\pi }{2}+\theta }(x)_{i\frac{\pi }{2}-\theta }=\frac{\cosh
\theta +\sin \pi x}{\cosh \theta -\sin \pi x} \,,\]
we get 
\[
-\log \left[ (x)_{i\frac{\pi }{2}+\theta }(x)_{i\frac{\pi }{2}-\theta
}\right] \, \sim \, -4\sin \pi x\, \mathrm{e}^{-\left| \theta \right|
}+O\left( \mathrm{e}^{-2\left| \theta \right| }\right)\ . 
\]
Note that \( k_{0} \) and \( A \) in (\ref{scatt_asymp}) can be calculated
additively from the asymptotics of the contribution of a single \( (x)\) block above.
As a result, \( k_{0}=0 \) and so the plateau eqn. (\ref{plateau_eqn}) has
no solution in the sinh-Gordon regime \( \lambda <0 \), thus the analytic continuation
described above cannot be avoided. Putting the ingredients together,
the boundary energy in sinh-Gordon theory takes the form
\begin{equation}
\label{sinh_benergy}
E^{\mathrm{shG}}(\eta ,\vartheta )=-\frac{m}{2\sin \frac{\pi }{\lambda
}}\left( \cos \left( \frac{\eta }{\lambda }\right) +\cosh \left(
\frac{\vartheta }{\lambda }\right) -\frac{1}{2}\cos \left( \frac{\pi
}{2\lambda }\right) +\frac{1}{2}\sin \left( \frac{\pi }{2\lambda
}\right) -\frac{1}{2}\right)\ . 
\end{equation}
It is now easy to recover Zamolodchikov's formula (\ref{boundary_energy}) for the boundary
energy in sine-Gordon theory.

As an immediate check on this calculation, we wish to note that for \(
\lambda =\frac{3}{2}\)
the \( S \)-matrix (\ref{s_matrix}) is identical to that of the scaling Lee-Yang
model, and the reflection factors of the scaling Lee-Yang model corresponding
to integrable boundary conditions are reproduced by specifying some complex values
for \( \eta  \) and \( \vartheta  \). It can be easily checked that the formula
(\ref{sinh_benergy}) reproduces correctly the results of Dorey et al. \cite{dptw}.

\subsection{Special cases}

Since we obtained the boundary energy of sine-Gordon/sinh-Gordon
theory under some non trivial assumptions we check the results in
some known cases.

\subsubsection{Dirichlet boundary conditions}

Dirichlet boundary conditions correspond to the limit \( \mu \, \rightarrow \, \infty  \)
in (\ref{pCFT_action}), which leads to\[
\Phi (0,\, t)=\phi _{0}\, \bmod \, \frac{2\pi }{\beta }\ .\]
 The reflection factor of the first breather can be obtained as the \( \vartheta \, \rightarrow \, \infty  \)
limit of (\ref{b1_refl}):
\[
R^{(1)}(\theta )=\frac{\left( \frac{1}{2}\right) _{\theta }\left(
\frac{1}{2\lambda }+1\right) _{\theta }}{\left( \frac{1}{2\lambda
}+\frac{3}{2}\right) _{\theta }}\frac{\left( \frac{\eta }{\pi \lambda
}-\frac{1}{2}\right) _{\theta }}{\left( \frac{\eta }{\pi \lambda
}+\frac{1}{2}\right) _{\theta }}\ .\]
The derivation of the previous subsection then gives the boundary energy 
\begin{equation}
\label{ebnd_dirichlet}
E_D(\eta )=-\frac{M}{2\cos \frac{\pi }{2\lambda }}\left( \cos \left( \frac{\eta }{\lambda }\right) -\frac{1}{2}\cos \left( \frac{\pi }{2\lambda }\right) +\frac{1}{2}\sin \left( \frac{\pi }{2\lambda }\right) -\frac{1}{2}\right)\,, 
\end{equation}
which is exactly identical to the formula obtained by Leclair et al. in \cite{leclair}.
The parameter \( \eta  \) is related to \( \phi _{0} \) in the following way\[
\eta =\pi \left( \lambda +1\right) \frac{\beta \phi _{0}}{2\pi }\,,\]
which was conjectured by Ghoshal and Zamolodchikov \cite{ghoshal}, and is a
straightforward consequence of eqns. (\ref{UV_IR_relation}) as well.

Note that $E_D(\eta)$ cannot be obtained as the
\(\vartheta \rightarrow\infty \) limit of the general boundary energy 
eqn. (\ref{boundary_energy}). The reason is clear: the boundary potential
is normalized in different ways in the two cases: classically to
obtain finite energy in the Dirichlet limit one has to add $M_0$ to
the general  
$-M_0\cos\left(\frac{\beta}{2}(\Phi -\phi_0)\right)$ boundary potential.
Clearly in the quantum case, when the boundary vertex operator has a
non trivial dimension, we can not simply subtract $\tilde{\mu}$ from
$E(\eta ,\vartheta)$. Since the quantity we subtract must have the
dimension of mass and should depend on $\tilde{\mu}$, it must be
proportional to
$\tilde{\mu}^{1/(1-h_\beta)}=\tilde{\mu}^{\lambda/(\lambda +1)}$. The
question is whether we can make this subtraction such that in the  
$\vartheta\rightarrow\infty$ limit the leading term cancels and 
the constant terms just reproduce
$E_D(\eta )$. The UV-IR relations, eqn. (\ref{UV_IR}-\ref{mu_crit})
guarantee, that
\[
\tilde{\mu}\rightarrow \frac{\mu_{\rm
crit}}{2}\exp\left(\frac{\vartheta}{\lambda +1}\right)\left( 1+
\exp\left(-\frac{2\vartheta}{\lambda +1}\right)\cos(\frac{2\eta}{\lambda
+1})+{\cal O}\exp\left(-\frac{4\vartheta}{\lambda +1}\right)
\right)\quad {\rm as}\quad \vartheta\rightarrow\infty  \ .
\]
Thus, upon using the bulk mass gap relation (cf. eqn. (\ref{kapa})),
$\tilde{\mu}^{\lambda/(\lambda +1)}$ becomes proportional to \(
Me^{\vartheta /\lambda }\) up to exponentially small terms for
$\vartheta\rightarrow\infty $. Therefore, if we subtract this term
with an appropriate coefficient then in the Dirichlet limit the
surviving constant terms exactly reproduce (\ref{ebnd_dirichlet}).

\subsubsection{The first excited state}

It was noted in \cite{mattsson} (for Dirichlet boundary condition)
and in \cite{bajnok1} (for the general case) that continuing
analytically
\[
\eta\,\rightarrow\,\pi(\lambda+1)-\eta
\]
the role of the boundary ground state \( |\rangle \) and the boundary
first excited state \( |0 \rangle \) are interchanged. Therefore we
can calculate the energy difference between these two states from the
formula for the boundary energy, eqn. (\ref{boundary_energy}). The
result is 
\[
E(\pi(\lambda+1)-\eta,\vartheta)-E(\eta,\vartheta)=
M\cos\left(\frac{\eta}{\lambda}-\frac{\pi}{2\lambda}\right)
\]
which exactly equals the prediction of the bootstrap, i.e. 
\[
E_{ |0 \rangle }-E_{ |\rangle }=
M\cos\nu_0
\]
that follows from eqn. (\ref{excited_energy}).

\subsection{UV-IR relations and vacuum expectation values (VEVs)}

As it is well known in the bulk sine-Gordon theory there is a relation
among the following three exactly calculable quantities: the ground
state energy density, the dimensionless constant entering the mass gap
relation connecting the UV and IR parameters, and the VEV of the
exponential field \(\langle e^{i\beta\Phi (x)}\rangle\) \cite{LZ}.
This relation is such that knowing any two of these quantities
determines the third one. It generalizes to sine-Gordon theory
with boundaries, where it connects the boundary energy, the UV-IR
relations (\ref{UV_IR}-\ref{mu_crit}), and the VEV of the boundary
field \(\langle e^{i\frac{\beta}{2}\Phi (0)}\rangle\) in a similar
way. As the VEV of the boundary operators has been determined by
Fateev, Zamolodchikov and Zamolodchikov (FZZ), we can use it to show
that the UV-IR relations, (\ref{UV_IR}-\ref{mu_crit}) and the boundary
energy, (\ref{boundary_energy}), are indeed consistent with the VEVs
given in \cite{FZZ}. For simplicity we consider only the special case
when $\phi_0=0$, as this case already illustrates the point. (More
precisely the condition $\phi_0=0$ can be satisfied in two different
ways \cite{bajnok1}: either by $\vartheta =0$ or by $\eta =0$, and we
consider the former possibility).

Writing the functional integral representation of the partition
function $Z_{ab}={\rm Tr}e^{-RH_{ab}(L)}$ on a cylinder of length $R$
and circumference $L$ with boundary states $a$ and $b$ on the boundary
circles and considering the $R\rightarrow\infty$ limit (when
$Z_{ab}\sim e^{-RE_{ab}(L)}$) one readily derives that in this limit
the ground state energy $E_{aa}$ satisfies 
\be\label{be1}
\frac{\partial E_{aa}}{\partial\tilde{\mu}}=-\langle
e^{i\frac{\beta}{2}\Phi (0)}\rangle \equiv - G(\beta,\tilde{\mu}).
\ee 
(In writing this equation we assumed that
$G(\beta,\tilde{\mu})=G(-\beta,\tilde{\mu})$). Since for $\vartheta
=0$ the ground state energy depends on $\tilde{\mu}$ only through the
$\eta $ parameter appearing in the boundary energy, eqn. (\ref{be1})
actually determines the dependence of $\eta $ on
$\tilde{\mu}$. Furthermore, both sides of (\ref{be1}) can be integrated
to obtain the following expression for the boundary energy
\be\label{be2} E(\eta )=-\int d\tilde{\mu} G(\beta,\tilde{\mu})\,.
\ee 
What we show below is that using the FZZ expression for
$G(\beta,\tilde{\mu})$ on the r.h.s. gives (\ref{boundary_energy}) for
the boundary energy only if (\ref{UV_IR}-\ref{mu_crit}) hold.

The expression given in \cite{FZZ} for $G(\beta,\tilde{\mu})$ depends
on $\tilde{\mu}$ through a parameter $z$, which, for $\phi_0=0$, we
take to be pure imaginary $z=iZ$ ($Z$ real):
\be\label{wbe}
\cos^2(\pi
Z)=\frac{\tilde{\mu}^2}{2\mu}\sin\left(\frac{\beta^2}{8}\right)\,.
\ee
Then
\[
G(\beta,\tilde{\mu})=\left(\frac{\pi\mu\Gamma
(\frac{\lambda}{\lambda +1})}{2\Gamma (\frac{1}{\lambda
+1})}\right)^{\frac{1}{2\lambda}}g_0(\beta )g_S(\beta ,Z)\,,
\]
where $g_0$ and $g_S$ are given by the integral
representations\footnote{Note that the integral for $\log g_S$
contains a factor of $1/2$ compared to the expression in \cite{FZZ}
even after accounting for the difference between the parameters of
this paper and of \cite{FZZ}. Without the inclusion of this factor it
would be impossible to obtain the correct $\eta$ dependence of the
boundary energy as in eqn. (\ref{boundary_energy}). The fact that this
factor should be present was later confirmed to us by
Al.B. Zamolodchikov in a private discussion.}:
\[
\log g_0(\beta )=\int\limits^\infty_0\frac{dt}{t}\left[
\frac{2\sinh(t/(\lambda +1)}{\sinh (t)\sinh (t\lambda /(\lambda +1))}    
\left( e^{t\lambda /(2\lambda +2)}\cosh \left(\frac{t}{2}\right)\cosh \left(\frac{t}{2\lambda
+2}\right)-1\right)-\frac{e^{-t}}{\lambda +1}\right]\,,
\]
\[
\log g_S(\beta ,Z)=-\int\limits^\infty_0\frac{dt}{t}
\frac{2\sinh(t/(\lambda +1)\sinh^2(Zt)}{\sinh (t)\sinh (t\lambda
/(\lambda +1))}\,.
\]    
The integrals appearing here can be computed analytically after some
efforts. Finally, expressing $\mu $ in terms of the soliton mass $M$
via (\ref{kapa}), and converting the integral over $\tilde{\mu}$ to an
integral over $\pi Z$ by using (\ref{wbe}), after some algebra one
finds
\[
-\int d\tilde{\mu}
G(\beta,\tilde{\mu})=-\frac{M}{2\cos\left(\frac{\pi}{2\lambda}\right)}
\cos\left(\frac{Z\pi (\lambda +1)}{\lambda}\right)+f(\lambda)\ .
\]
This agrees with the boundary energy, (\ref{boundary_energy}), if
$Z\pi=\frac{\eta}{\lambda +1}$, i.e. when eqn. (\ref{wbe}) becomes
identical to (the $\vartheta =0$ case of) (\ref{UV_IR}-\ref{mu_crit}).

\section{\label{sec:general_TCSA} TCSA: general integrable boundary condition}

\subsection{TCSA for the boundary sine-Gordon model}

First we describe the Hamiltonian of boundary sine-Gordon model (BSG) living
on the line segment \( 0\leq x\leq L \) as that of a bulk and boundary perturbed
free boson with suitable boundary conditions. This is the starting point of
the TCSA analysis.

The basic idea of TCSA is to describe certain \( 2d \) models in
finite volume as relevant perturbations of their ultraviolet limiting
CFT-s \cite{YZ}. If we consider boundary field theories, then the
CFT-s in the ultraviolet are in fact boundary CFT-s. The use of TCSA
to investigate boundary theories was advocated in \cite{dptw,ger}.

As the bulk SG model can be successfully described in TCSA as a perturbation of the
\( c=1 \) free boson \cite{frt}, it is natural to expect that the various
BSG models are appropriate perturbations of \( c=1 \) theories with Neumann
or Dirichlet boundary conditions. Therefore we take the strip \( 0\leq x\leq L \)
and consider the following perturbations of the free boson, as described in detail in
\cite{neumann}: 
\begin{eqnarray*}
S & = & \displaystyle\int ^{\infty }_{-\infty }\int _{0}^{L}\left( \frac{1}{2}\partial _{\mu }\Phi \partial ^{\mu }\Phi +\mu \cos (\beta \Phi )\right) dxdt+\\
 &  & +\int ^{\infty }_{-\infty }\left( \tilde{\mu }_{0}\cos \left( \frac{\beta }{2}(\Phi _{B}-\phi _{0})\right) +\tilde{\mu }_{L}\cos \left( \frac{\beta }{2}(\Phi _{B}-\phi _{L})\right) \right) dt\, \, .
\end{eqnarray*}

Here, for finite \( \tilde{\mu } \)'s, Neumann boundary conditions are imposed
in the underlying \( c=1 \) theory on the boundaries, while if any of the \( \tilde{\mu } \)
-s is infinite then the corresponding term is absent and the boundary condition
in the underlying conformal theory on that boundary is Dirichlet. 
The Hamiltonian of the system can be rewritten in terms of the variables associated to the plane
using the map \( (x,it)=\xi \, \rightarrow \, z=e^{i\frac{\pi }{L}\xi } \): 
\begin{eqnarray}
H= & H_{CFT}-\displaystyle\frac{\mu }{2}\left( \frac{\pi }{L}\right) ^{2h_{\beta }-1}\displaystyle\int _{0}^{\pi }\left( V_{\beta }(e^{i\theta },e^{-i\theta })+V_{-\beta }(e^{i\theta },e^{-i\theta })\right) d\theta - & \nonumber \\
 & \displaystyle\frac{\tilde{\mu }_{0}}{2}\left( \frac{\pi }{L}\right) ^{h_{\beta }}\left( e^{-i\frac{\beta }{2}\phi _{0}}\Psi _{\frac{\beta }{2}}(1)+e^{i\frac{\beta }{2}\phi _{0}}\Psi _{-\frac{\beta }{2}}(1)\right) - & \nonumber \\
 & \displaystyle\frac{\tilde{\mu }_{L}}{2}\left( \frac{\pi }{L}\right) ^{h_{\beta }}\left( e^{-i\frac{\beta }{2}\phi _{L}}\Psi _{\frac{\beta }{2}}(-1)+e^{i\frac{\beta }{2}\phi _{L}}\Psi _{-\frac{\beta }{2}}(-1)\right) \, \, .\label{tcsa_ham} 
\end{eqnarray}
Here \( V_{\beta }(z,\bar{z})=n(z,\bar{z}):\, e^{i\beta \Phi (z,\bar{z})}: \)
and \( \Psi _{\frac{\beta }{2}}(y)=:\, e^{i\frac{\beta }{2}\Phi (y,y)}: \)
are the bulk and boundary vertex operators and the normal ordering coefficient
\( n(z,\bar{z}) \) depends on the boundary conditions chosen \cite{neumann}. 

Now the computation of the matrix elements of the bulk and boundary
vertex operators 
\( V_{\pm \beta } \) and \( \Psi _{\pm \beta /2} \)
(with conformal dimension \( h_{\beta }=\frac{\beta ^{2}}{8\pi } \))
between the vectors of the appropriate conform Hilbert spaces is
straightforward and the integrals can also be calculated
explicitly. The TCSA method consists of truncating the Hilbert space
at a certain conformal energy level \( E_{\mathrm{cut}} \) (which is
nothing but the eigenvalue of the zeroth Virasoro generator) and
diagonalizing the Hamiltonian numerically.

It is important to realize that one has to write separate TCSA
programs for checking the Dirichlet limit and the general two
parameter case. In the Dirichlet case there are no relevant operators
on the boundary, thus both $\tilde{\mu}_0$ and $\tilde{\mu}_L$ must be
set to zero, and we can have $\tilde{\mu}$-s different from zero only
if we perturb a CFT with Neumann boundary condition. Therefore we
investigate the general two parameter boundary sine-Gordon theory by 
describing it as an appropriately perturbed $c=1$ CFT with Neumann
boundary conditions at both ends.     
The Hilbert
spaces of the $c=1$ CFT-s with Dirichlet or Neumann boundary
conditions at the two ends are rather different: while in the former case it
basically consists of the vacuum module only, in the latter it is the
direct sum of modules built on the highest weight vectors carrying
the allowed values of the field momentum. 

Let us investigate the general two parameter BSG first. Then the
simplest choice (i.e. the one resulting in the least complex spectrum
which is enough to compare to the predictions) is to
switch on the boundary perturbation only at one end of the strip.  
The TCSA Hamiltonian for BSG with Neumann boundary condition at one end and perturbed
Neumann condition, (\ref{hatfel}), at the other, is obtained from
(\ref{tcsa_ham}) 
by setting \( \tilde{\mu }_{L}=0 \), \( \tilde{\mu }_{0}\equiv 
\tilde{\mu }\ne 0 \) .
The spectrum of vertex operators in this case is \( V_{\frac{n}{r}}(z,\bar{z}) \)
and \( \Psi _{\frac{m}{r}}(y) \), where \( r \) is the compactification radius
of the free boson of the $c=1$ theory in the UV, and \( n \), $m$ are
integers. 
These fields are primary under
the chiral algebra \( \widehat{U(1)} \) (i.e. \( U(1) \) affine Lie algebra).
However the compactification radius must be chosen so that 
 both  \( V_{\pm \beta } \) and \( \Psi _{\pm \frac{\beta }{2}} \)be
in the spectrum:\footnote{The $\sqrt{4\pi}$
has its origin in the different normalizations of the SG scalar field
$\Phi$ and the $c=1$ CFT one.}  $r=2\sqrt{4\pi}/\beta =2r_0$. Then \( V_{\pm \beta } \) are
represented as \( V_{\pm\frac{2}{r}}\) while \( \Psi _{\pm \frac{\beta
}{2}} \) as \( \Psi _{\pm\frac{1}{r}}\).  
In other words we have to consider
the boundary perturbation of the \( 2 \)-folded sine-Gordon model in the sense of \cite{kfold}. 

We choose our units in terms of the soliton mass \( M \). The bulk coupling
\( \mu  \) is related to \( M \) by \begin{equation}\label{kapa}
\mu =\kappa (\beta )M^{2-2h_{\beta }},\qquad \qquad h_{\beta }=\frac{\beta ^{2}}{8\pi },
\end{equation}
 where \( \kappa (\beta ) \) is a dimensionless constant. In the bulk SG, from
TBA considerations, the exact form of \( \kappa (\beta ) \) was obtained in
\cite{massgap}, and we use the same form also here in BSG. Once we expressed
\( \mu  \) and used the UV-IR relation
(\ref{UV_IR_relation},\ref{mu_crit}) to rewrite 
$\tilde{\mu }\exp\left(\pm i\frac{\beta}{2}\phi_0\right)$ in terms of
the IR parameters,  
 the Hamiltonian can be made dimensionless \( h=H/M \), depending only
on the dimensionless volume  \( l=ML \), the coupling constant \( \beta \)  
and \( \eta ,\vartheta  \). 
We compare the predictions on the spectrum, ground state energy etc. 
of the general two parameter BSG model to
the truncated spectrum of this Hamiltonian.

\subsection{Finite size corrections from scattering theory}

Here we briefly recall the method to calculate the finite size corrections
for large volumes (\( l\gg 1 \)) from the knowledge of the bulk \( S \)-matrices
and boundary reflection factors. To simplify the presentation, let us consider
a single scalar particle of mass \( m \) with reflection factors \( R_{a}\left( \theta \right)  \)
and \( R_{b}(\theta ) \) on the boundaries at \( x=0 \) and \( x=L \) respectively.
Then the energy as a function of the volume can be obtained by solving the Bethe-Yang
equation\begin{equation}
\label{bethe_yang_eqn}
mL\sinh \theta -i\log R_{a}\left( \theta \right) -i\log R_{b}\left( \theta \right) =2\pi I
\end{equation}
for \( \theta  \), where \( I \) is an integer (half integer) quantum number (corresponding
to quantization of momentum in finite volume). From the solution
$\theta(L)$ of (\ref{bethe_yang_eqn}) the energy with respect to the
state with no particles is obtained as \begin{equation}
\label{bethe_yang_energy}
E\left( L \right) -E_{0}^{ab}\left( L \right) =m\cosh \theta\left( L\right) .
\end{equation}
Eqn. (\ref{bethe_yang_eqn}-\ref{bethe_yang_energy}) can also be used
to give the \( (E(L),L)\) \lq Bethe-Yang line' in a parametric form.
When \( I=0 \), eqn. (\ref{bethe_yang_eqn}) may have solutions
corresponding to purely imaginary \( \theta \), which may (in turn)
correspond to boundary excited states obtained from the particle
binding to one of the walls, cf. \cite{neumann} for details.

\subsection{Results}

In the TCSA for the general two parameter case the number of states
with conformal energies below $E_{\rm cut}$ depends very sensitively
on the coupling constant $\beta$ (compactification radius $r$), since
the Hilbert space of the conformal free boson with Neumann boundary
conditions is the direct sum of modules corresponding to the various
momenta, which are integer multiple of $1/r$. Therefore it is not a
surprise that in the range $r_0\ge 3/2$, where TCSA is expected to
converge, there are so many states even for moderate $E_{\rm cut}$-s,
that the time needed for diagonalizing $H$ practically makes it
impossible to proceed.

We overcome this difficulty partly by considering first only models on
a ``special line'' in the parameter space described by $\phi_0=0$ or
$\vartheta =0$. As pointed out in \cite{bajnok1}, the models on this
line admit the $\Phi\mapsto -\Phi $ \lq charge conjugation' symmetry
as a result of the equality $P^+=P^-\equiv P$.  As a consequence in
these models there are two sectors, namely the even and the odd ones.
It is straightforward to implement the projection onto the even and
odd sectors in the conformal Hilbert spaces used in TCSA. This
projection has two beneficial effects: on the one hand it effectively
halves the number of states below $E_{\rm cut}$\footnote{In our
numerical studies of these models $E_{\rm cut}$ varied between 15 and
18 and this resulted in $3\times 10^3$ - $5\times 10^3$ conformal
states per {\it sectors}.}, thus it drastically reduces the time
needed to obtain the complete TCSA spectrum, and on the other the
separate spectra of the even and odd sectors are less complex and
therefore easier to study than the combined one. Furthermore, the
spectrum of boundary states in the most general case depends only on
$\eta$ \cite{bajnok1}, and so our considerations can be restricted to 
$\vartheta =0$ without any loss of generality in this respect.
  
\subsubsection{Boundary energy}

\begin{figure}
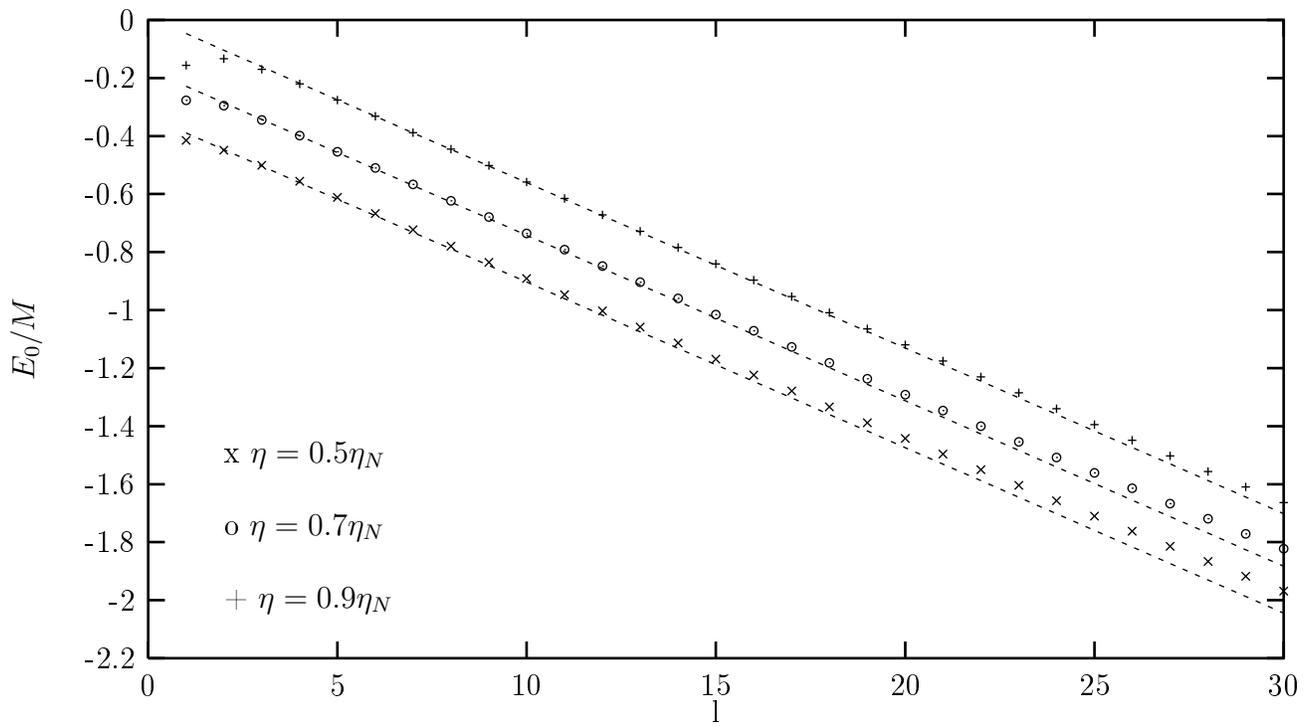

\psfrag{etan}{$\eta_N$}
\psfrag{eta}{$\eta$}
\centering
\include{eplo.psltx}
\caption{Ground state energy versus $l$ in three BSG models with
$r_0=\sqrt{4\pi}/\beta=2$ and $\vartheta =0$.}
\label{fig:eplo}
\end{figure}

\begin{figure}
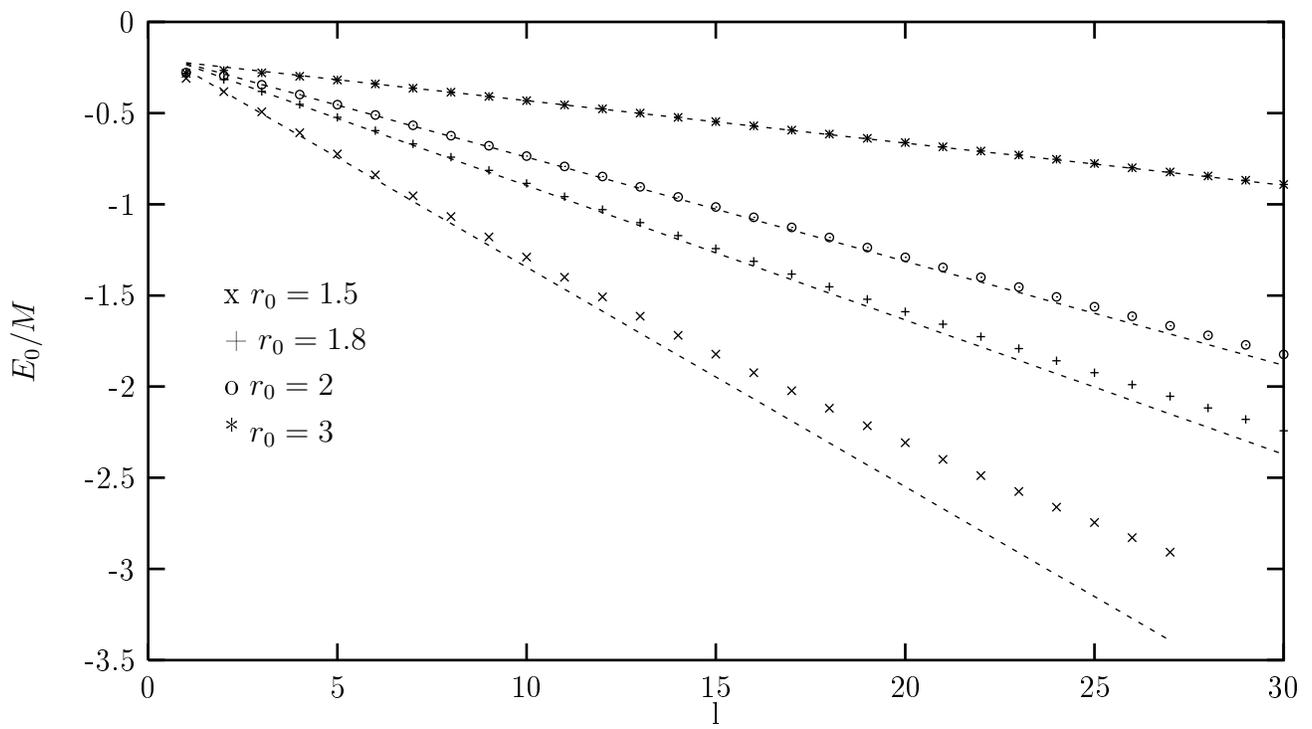

\centering
\include{rplo.psltx}
\caption{Ground state energy versus $l$ in four BSG models with
$\eta=0.7\eta_N$ and $\vartheta =0$. 
 }
\label{fig:rplo}
\end{figure}

First we investigate the ground state energy of these models to check
the predictions of the BSG model. Since at one end of the strip we
imposed ordinary Neumann boundary condition and switched on the
boundary perturbation only at the other end, the ground state energy
(in units of the soliton mass) for large enough $L$-s should depend on
the dimensionless volume $l=ML$ as \be\label{komalap}
\frac{E}{M}(l)=-\frac{l}{4}\tan(\frac{\pi}{2\lambda
})+\frac{E(\eta_N,0)}{M} +\frac{E(\eta
,0)}{M}+O\left(\mathrm{e}^{-l}\right)\,, \ee where $E(\eta ,\vartheta
)$ is the boundary energy, eqn. (\ref{boundary_energy}), and
$\eta_N=\frac{\pi}{2}(1+\lambda)$ is the $\eta$ parameter of the
Neumann boundary \cite{ghoshal}. This prediction is compared to the
TCSA data on Fig.s(\ref{fig:eplo}-\ref{fig:rplo}), where the dashed
lines are given by eqn. (\ref{komalap}). The agreement between the
predictions and the data is so good that in the interval $5\le l\le
15$ the bulk energy constant and the sum of boundary
energies can be measured with a reasonable accuracy.

In our earlier paper \cite{neumann}
when numerically investigating the ground state energy of the BSG model with
Neumann boundary condition at both ends we made a conjecture that 
\[ E(\eta_N,0)=-E_D(0)\]
holds. ($E_D$ is the boundary energy of the BSG model with Dirichlet
boundary condition, eqn. (\ref{ebnd_dirichlet})). Clearly the exact
expressions eqn. (\ref{boundary_energy}) and eqn. (\ref{ebnd_dirichlet})
do not satisfy this, but the violation of this relation is
practically undetectable (using numerical methods) in the
$\lambda $ range investigated in \cite{neumann}. 


We also checked the $\vartheta $ dependence of the boundary energy
$E(\eta ,\vartheta )$, eqn. (\ref{boundary_energy}): the rapid growth
in the number of states, caused by the absence of the two sectors, can
be compensated by going to a sufficiently attractive value of $\lambda
$ ($\lambda =17$) where TCSA is known to converge faster. In this case
the choice $E_{\rm cut}=13$ resulted in 4147 conformal states and we
could measure the sum of the two boundary energies fitting the volume
dependence of the ground state energy by a straight line in the range
$6\le l\le 17$, the results are collected in table \ref{thetadep}.

\begin{table}
{\centering \begin{tabular}{|c|c|c|}
\hline 
\( \vartheta  \)&
\( E(\eta_N,0)+E(\eta ,\vartheta ) \) (predicted)&
\( E(\eta_N,0)+E(\eta ,\vartheta ) \) (TCSA)\\ 
\hline 
\( 5 \)&
\( -0.22259 \)&
\( -0.226959 \)\\
\hline 
\( 10 \)&
\(-0.29012  \)&
\(-0.29986  \)\\
\hline 
\end{tabular}\par}

\caption{\label{thetadep} Boundary energies (in units of soliton mass)
of two BSG models with
$\lambda =17$ and $\eta =0.7\eta_N$ 
as measured from TCSA}
\end{table}
To sum up, we showed that the prediction eqn. (\ref{boundary_energy})
for the boundary energy of the general two parameter boundary
sine-Gordon model is in perfect agreement with the TCSA data. 
This agreement indirectly confirms also the UV-IR relations, 
eqn. (\ref{UV_IR}-\ref{mu_crit}), since they were built into the TCSA
program. The case
of Dirichlet boundary conditions is investigated in the next section.

\subsubsection{Reflection factors and the spectrum of excited states}

We compare the reflection factors and the spectrum of excited states
to the TCSA data in the case of models with $\phi_0=0$ (which is
realized here as $\vartheta =0$).  The bulk breathers naturally
belong to one of the sectors, as the ${\bf C}$ parity of the $n$-th
breather is $(-1)^n$. However, since solitons and anti solitons can
reflect into themselves as well as into their charge conjugate
partners, solitonic one particle states (i.e. states, whose energy and
momentum are related by $E=\sqrt{P^2+M^2}$ where $M$ is the soliton
mass) are present in both sectors.

To associate the various boundary bound states to the two sectors we
have to determine the ${\rm C}$ parity of the poles $\nu_n$ and $w_m$
in the soliton/antisoliton reflection factors. As in the even/odd
sectors the reflection factors are given by $P\pm Q$ (where $P\equiv
P^+=P^-$ for $\vartheta =0$), the possible cancellation between the
zeroes of $P_0\pm Q_0$ and the poles of $\sigma(\eta ,u)$ have to be
investigated. The outcome is that the poles at $\nu_{2k}$ and $w_{2k}$
($k=0,1,2,..$) appear in $P+Q$ (i.e. the corresponding bound states
are in the even sector), while the poles $\nu_{2k+1}$, $w_{2k+1}$
appear in $P-Q$ (i.e. the corresponding bound states are in the odd
sector).

We analyzed the appearance of boundary bound states in the TCSA
spectra of a number of BSG models. The results are illustrated on the
example of a model when $\lambda =7$ and $\eta =0.9\ \eta_N$. For
these values of the parameters the sequence of $\nu_n$-s and $w_m$-s
in the physical strip is 
\be \nu_0>w_1>\nu_1>w_2>\nu_2>w_3>\nu_3\,.
\ee 
Therefore in the even sector we expect the following low lying
bound states (i.e. ones with not more then three
labels\footnote{States having more labels are heavier thus they
correspond to higher TCSA lines.}): 
\be \vert 0\rangle ,\quad \vert
2\rangle ,\quad \vert 0,2\rangle ,\quad \vert 1,3\rangle ,\quad \vert
0,1,1\rangle ,\quad \vert 0,1,3\rangle ,\quad \vert 1,2,3\rangle
,\quad \vert 2,3,3\rangle , \ee while in the odd sector \be \vert
1\rangle ,\quad \vert 3\rangle ,\quad \vert 1,2\rangle ,\quad \vert
2,3\rangle ,\quad \vert 0,1,2\rangle ,\quad \vert 1,2,2\rangle .  
\ee
Since at one end of the strip the unperturbed Neumann boundary
condition is imposed, the corresponding bound states are also expected
to appear in the TCSA spectrum. As described in
\cite{neumann}-\cite{bajnok1} for $\eta =\eta_N$ the $\nu_n$-s and the
$w_m$-s coincide and the bound states can be labeled by an increasing
sequence of positive integers $\vert n_1,...,n_k\rangle _N$ with
$n_k\le \lambda /2$. Therefore in the even sector there should be TCSA
lines corresponding to the 
\be \vert 2\rangle _N, \quad \vert
1,3\rangle _N, \quad \vert 1,2,3\rangle _N, \ee \lq Neumann bound
states', while in the odd one to \be \vert 1\rangle _N, \quad \vert
3\rangle _N, \quad \vert 1,2\rangle _N, \quad \vert 2,3\rangle _N.
\ee 
Finally there should be TCSA lines describing the
situation when both boundaries are in excited states with no
particle(s) moving between them, thus e.g. one expects a line in the
even (odd) sector that corresponds to $\vert 0\rangle\otimes\vert
2\rangle _N$ ($\vert 0\rangle\otimes\vert 1\rangle _N$).

\begin{figure}
{\centering \begin{tabular}{cc}
{\includegraphics{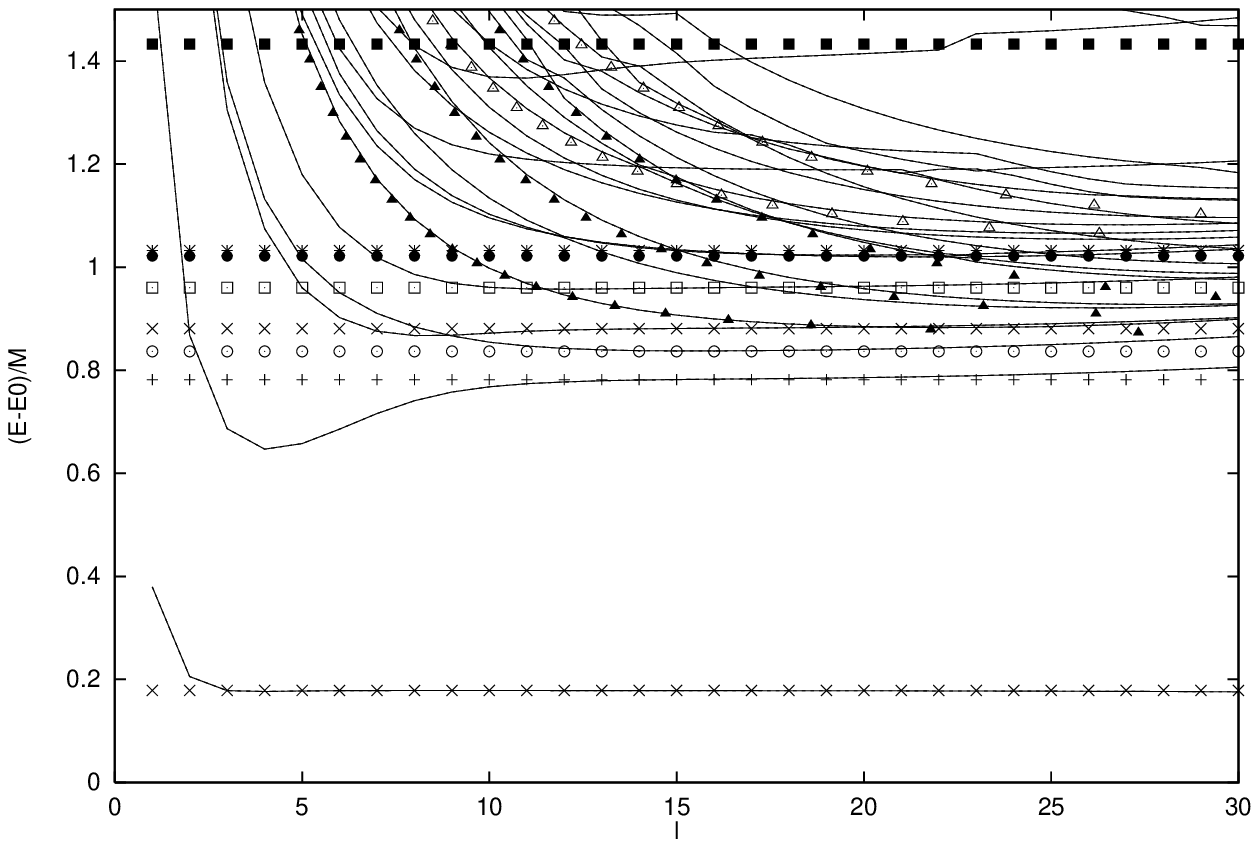}} \\
{\small The even sector: x denote the energy of $\vert 0\rangle $ and
$\vert 2\rangle $, $+$ that of $\vert 2\rangle _N$, $\circ$ of $\vert 0,2\rangle $,}\\{\small 
$\bullet $, the empty/full squares stand for $\vert 1\rangle
_N\otimes\vert 1\rangle $, $\vert 2\rangle
_N\otimes\vert 0\rangle $ and $\vert 1\rangle_N\otimes\vert 3\rangle $, $*$ for $\vert 0,1,1\rangle $,}
\\{\small the full/empty triangles are $B^2$ lines on ground
state/$\vert 0\rangle $ boundary.}\\
{\includegraphics{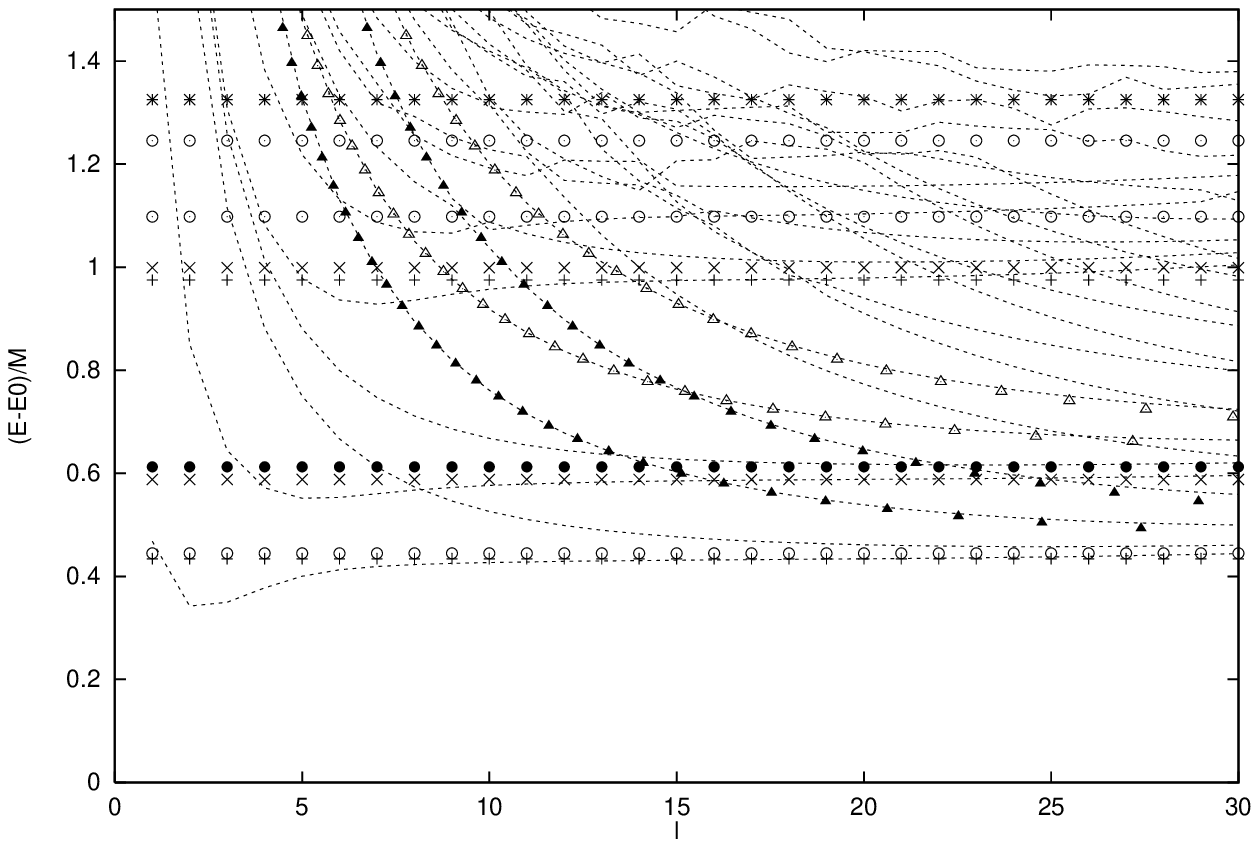}} \\
{\small  The odd sector: x stand for the energy of $\vert 1\rangle $,
$\vert 3\rangle $ ,
$+$ for $\vert 1\rangle_N$, $\vert 3\rangle_N$, $\bullet $ for $\vert 1\rangle
_N\otimes\vert 0\rangle $,}
\\{\small 
$\circ$ stand for $\vert 0,1\rangle $, $\vert 0,3\rangle $ and $\vert
1,2\rangle $, $*$ for $\vert 0,1,2\rangle $,}
\\{\small the full/empty triangles are $B^1$ lines on ground
state/$\vert 0\rangle $ boundary. }
\\
\end{tabular}\small \par}
\caption{TCSA data, boundary bound states and breather Bethe Yang
lines in the BSG model with $\lambda =7$ and $\eta=0.9\ \eta_N$.}\label{fig:spektra}
\end{figure}

We compare the predictions about these bound states to the TCSA data
on Fig.(\ref{fig:spektra}) where the
dimensionless energy levels above the ground state are plotted against
$l$. On both plots the continuous lines are the interpolated TCSA data
and the various symbols mark the data corresponding to the various
boundary bound states and Bethe-Yang 
lines\footnote{Some of the higher TCSA lines appear to have been broken, the
apparent turning points are in fact level crossings with the other
line not shown. This happens because our numerical routine, instead of
giving the eigenvalues of the Hamiltonian in increasing order at each
value of $l$, fixes their order at a particular small $l$ and follows them
 -- keeping their order -- according to some criteria as $l$ is changing to higher values.}. 

The two plots on Fig.(\ref{fig:spektra}) show in a convincing way that
the low lying boundary states indeed appear as predicted by the
bootstrap solution. (We show only those really low lying ones whose
identification is beyond any doubt; the higher lying ones may be lost
among the multitude of other TCSA lines). The reader's attention is
called to two relevant points: first there is no TCSA line that would
correspond to a $\vert 1,1\rangle $ bound state. The absence of this
state is explained in the bootstrap solution \cite{bajnok1} by a
Coleman-Thun diagram, that exists only if $w_1>\nu_1$. Second, both in
the even and in the odd sectors, there is evidence for the existence
of the lowest bound states with three labels. These states are
predicted in the bootstrap solution by the absence of any Coleman-Thun
diagrams when $\nu_{n_1}>w_{n_2}>\nu_{n_3}$ holds. These two findings
together give an indirect argument for the correctness of the boundary
Coleman-Thun mechanism. This is most welcome, as the theoretical
foundations of the boundary version of this mechanism are less solid
than that of the bulk one.

On the plots on Fig.(\ref{fig:spektra}) we also show 
in case of the lightest breathers $B^1$, $B^2$ 
the excellent
agreement between the TCSA data and the energy levels as predicted by
the Bethe-Yang equations
(\ref{bethe_yang_eqn},\ref{bethe_yang_energy}), using either the
ground state reflection factors (\ref{Rbr1}, \ref{Rbr2}) or the ones on the $\vert 0\rangle $
excited boundary (\ref{Rbe1}, \ref{Rbe2}). 
 (In the latter case
one has to take into account that now $k$ is odd, $b_0^n(\eta ,u)=1$,
and the energy above the ground state also contains the energy of $\vert 0\rangle $).

\section{\label{sec:dirichlet_TCSA} TCSA: Dirichlet boundary conditions}

For Dirichlet boundary conditions, the formula (\ref{tcsa_ham}) has to be changed:
the terms containing boundary perturbations must be omitted, since there are
no relevant boundary operators on a Dirichlet boundary. Furthermore, one must
quantize the \( c=1 \) free boson with Dirichlet boundary condition, which
preserves boundary conformal invariance as well as the Neumann one. The Hilbert
space is also changed, because there is a single vertex operator 
(the identity) living on the boundary, therefore it is essentially the same
as the vacuum module of the chiral algebra (which in this case is the \( \widehat{U(1)} \)
affine Lie algebra). In all numerical computations the truncation level was
\( E_{\mathrm{cut}}=22 \), which corresponds to \( 4508 \) vectors.

\subsection{Boundary energy}

\begin{figure}
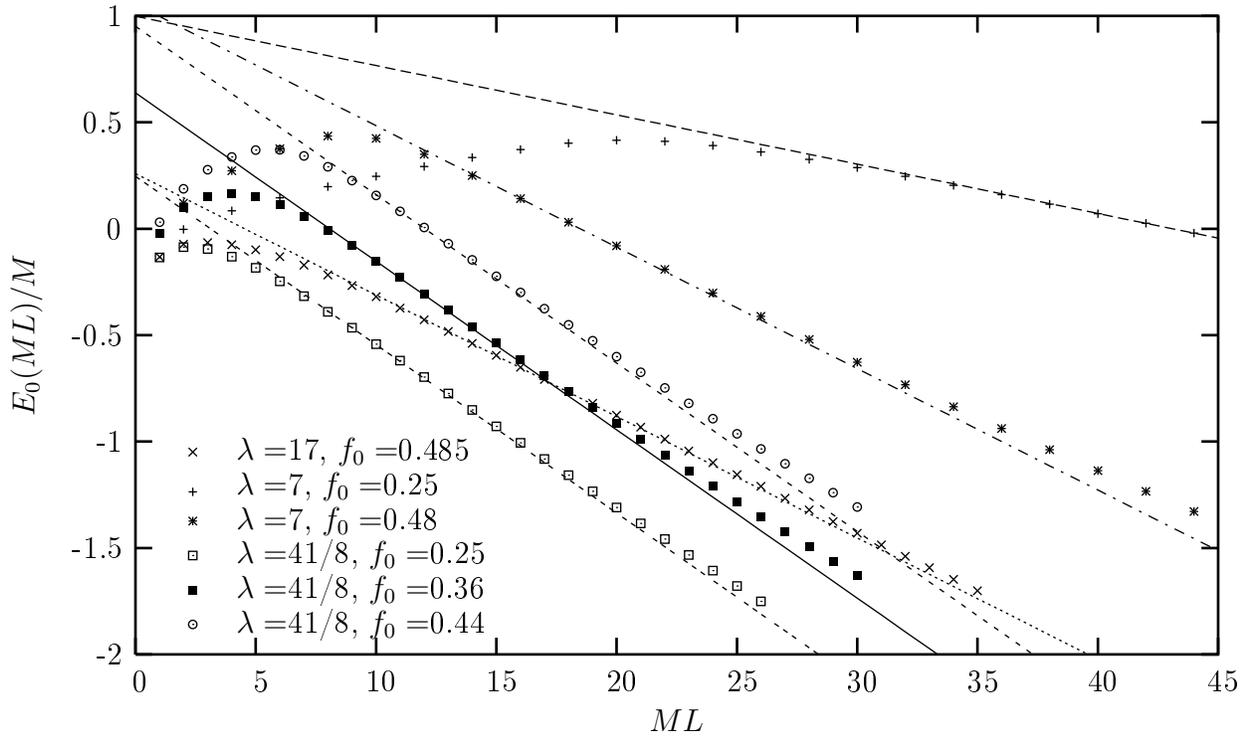

\psfrag{E0}{$E_0$}
\psfrag{la}{$\lambda$}
\include{ben.psltx}
\caption{\label{ben_Dir} Comparing the predicted bulk and boundary
energies to the TCSA data for Dirichlet boundary conditions. The dots
are the TCSA data for various values of $\lambda$ and $f_0=\frac{\beta\phi_0}{2\pi}$, while the
lines are their predicted asymptotic behaviour for large volume.}
\end{figure}

Here we summarize the agreement between the formula
(\ref{ebnd_dirichlet}), first derived in \cite{leclair} and TCSA with
Dirichlet boundary conditions. At both ends of the strip, identical
boundary conditions are imposed. In this case, it is easier to vary
the field value \( \phi_{0} \): the interaction needs to be calculated
for each given value of the sine-Gordon coupling parameter \( \lambda
\) only once. The agreement between the predicted values of the bulk and
boundary energy and the TCSA vacuum energy levels is illustrated on
Figure \ref{ben_Dir}, while numerical results are summarized in table
\ref{benergy_dir_table}.

\begin{table}
{\centering \begin{tabular}{|c|c|r|r|r|r|}
\hline 
\( \lambda  \)&
\( \frac{\beta \phi_{0}}{2\pi } \)&
\( E_{\mathrm{bulk}} \) (exact)&
\( E_{\mathrm{bulk}} \) (TCSA)&
\( E_{\mathrm{boundary}} \) (exact)&
\( E_{\mathrm{boundary}} \) (TCSA)\\
\hline 
\hline 
\( 31 \)&
\( 0 \)&
\( -0.01267857 \)&
\( -0.01267(2) \)&
\( -0.0259997 \)&
\( -0.026(17) \)\\
\hline 
\( 31 \)&
\( 0.2 \)&
\( -0.01267857 \)&
\( -0.0126(14) \)&
\( 0.1773231 \)&
\( 0.17(30) \)\\
\hline 
\( 31 \)&
\( 0.495 \)&
\( -0.01267857 \)&
\( -0.012(25) \)&
\( 1.009779 \)&
\( 0.97(75) \)\\
\hline 
\( 17 \)&
\( 0 \)&
\( -0.02316291 \)&
\( -0.0231(22) \)&
\( -0.0484739 \)&
\( -0.049(06) \)\\
\hline 
\( 17 \)&
\( 0.485 \)&
\( -0.02316291 \)&
\( -0.022(67) \)&
\( 0.998483 \)&
\( 0.97(78) \)\\
\hline 
\( 17 \)&
\( 0.5 \)&
\( -0.02316291 \)&
\( -0.022(69) \)&
\( 1.048474 \)&
\( 1.02(84) \)\\
\hline 
\( 7 \)&
\( 0.25 \)&
\( -0.05706087 \)&
\( -0.056(25) \)&
\( 0.259213 \)&
\( 0.24(88) \)\\
\hline 
\( 7 \)&
\( 0.48 \)&
\( -0.05706087 \)&
\( -0.055(62) \)&
\( 1.054646 \)&
\( 1.03(23) \)\\
\hline 
\( 41/8 \)&
\( 0.25 \)&
\( -0.07911730 \)&
\( -0.077(36) \)&
\( 0.2464426 \)&
\( 0.23(14) \)\\
\hline 
\( 41/8 \)&
\( 0.36 \)&
\( -0.07911730 \)&
\( -0.076(92) \)&
\( 0.6381842 \)&
\( 0.61(72) \)\\
\hline 
\( 41/8 \)&
\( 0.44 \)&
\( -0.07911730 \)&
\( -0.076(56) \)&
\(  0.9513045\)&
\(  0.92(52) \)\\
\hline 
\( 7/2 \)&
\( 0 \)&
\( -0.1203937 \)&
\( -0.118(22) \)&
\( -0.2957454 \)&
\( -0.30(11) \)\\
\hline 
\( 7/2 \)&
\( 0.3 \)&
\( -0.1203937 \)&
\( -0.114(69) \)&
\( 0.4241742 \)&
\( 0.39(37) \)\\
\hline 
\( 7/2 \)&
\( 0.42 \)&
\( -0.1203937 \)&
\( -0.11(34) \)&
\( 0.9532802 \)&
\( 0.91(12) \)\\
\hline 
\( 7/2 \)&
\( 0.5 \)&
\( -0.1203937 \)&
\( -0.11(18) \)&
\( 1.295745 \)&
\( 1.23(60) \)\\
\hline 
\end{tabular}\par}

\caption{\label{benergy_dir_table} Boundary energy for Dirichlet boundary conditions:
comparison to the TCSA data. The values for the boundary energy are for two
identical boundary conditions at both ends of the strip. Energies are given
in units of the soliton mass.}
\end{table}

\subsection{Reflection factors}

\begin{figure}
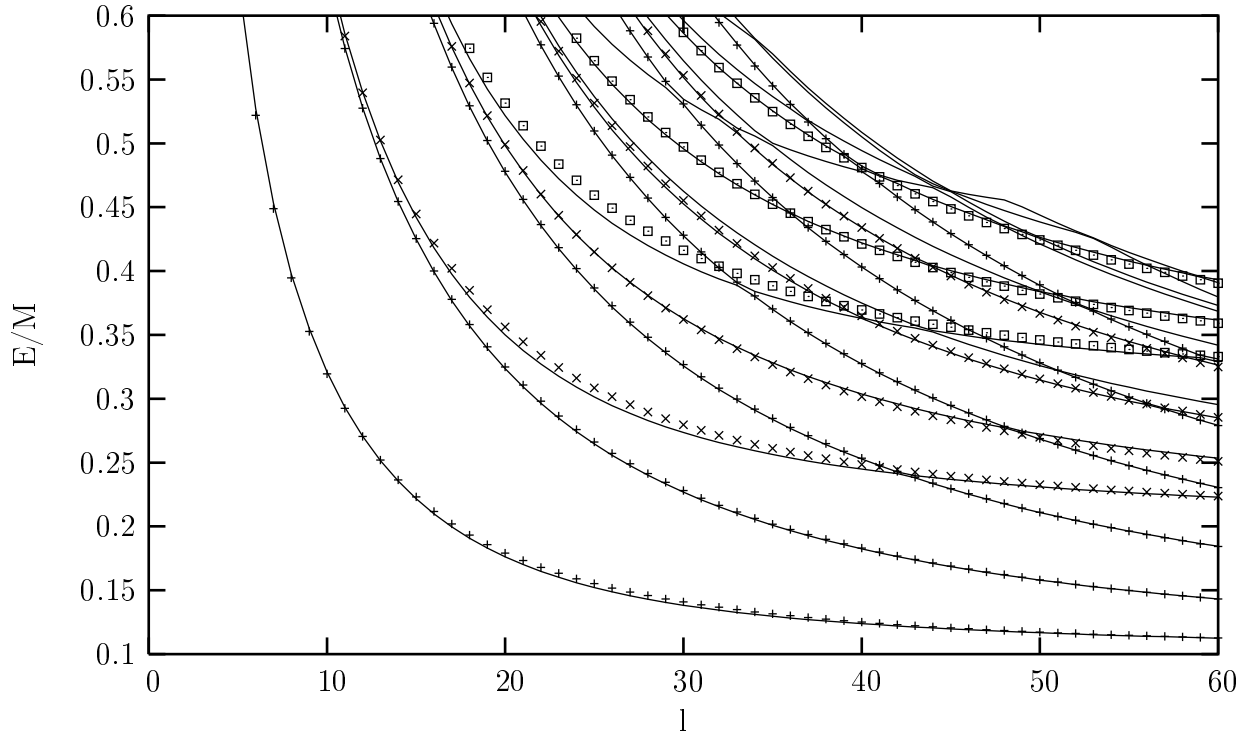

\include{refl_factors1.psltx}

\caption{\label{refl_factors1} Checking the reflection factors of
\protect\( B_{1}\protect \), \protect\( B_{2}\protect \) and
\protect\( B_{3}\protect \) for \protect\( \lambda =31\protect \) and
\protect\( \frac{\beta \phi_{0}}{2\pi }=0.2\protect \). The dots show
the one-particle energies predicted from the Bethe-Yang equations,
while the continuous lines are the TCSA results. All energies are
relative to the ground state and are in units of the soliton mass.}
\end{figure}

Using the Bethe-Yang equations (\ref{bethe_yang_eqn},\ref{bethe_yang_energy}),
we checked that the predictions for the energy levels from the ground state
reflection factors are in excellent agreement with the TCSA data. Figure \ref{refl_factors1}
is just an illustrative example; for all the other values of \( \lambda  \) and
\( \phi_{0} \) in Table \ref{benergy_dir_table} we had similar results. The
deviations are partly due to truncation effects, but partly signal the fact
that the Bethe-Yang equation only gives an approximate description of the finite
size corrections.

\begin{figure}
{\par\centering \resizebox*{1\columnwidth}{!}{\includegraphics{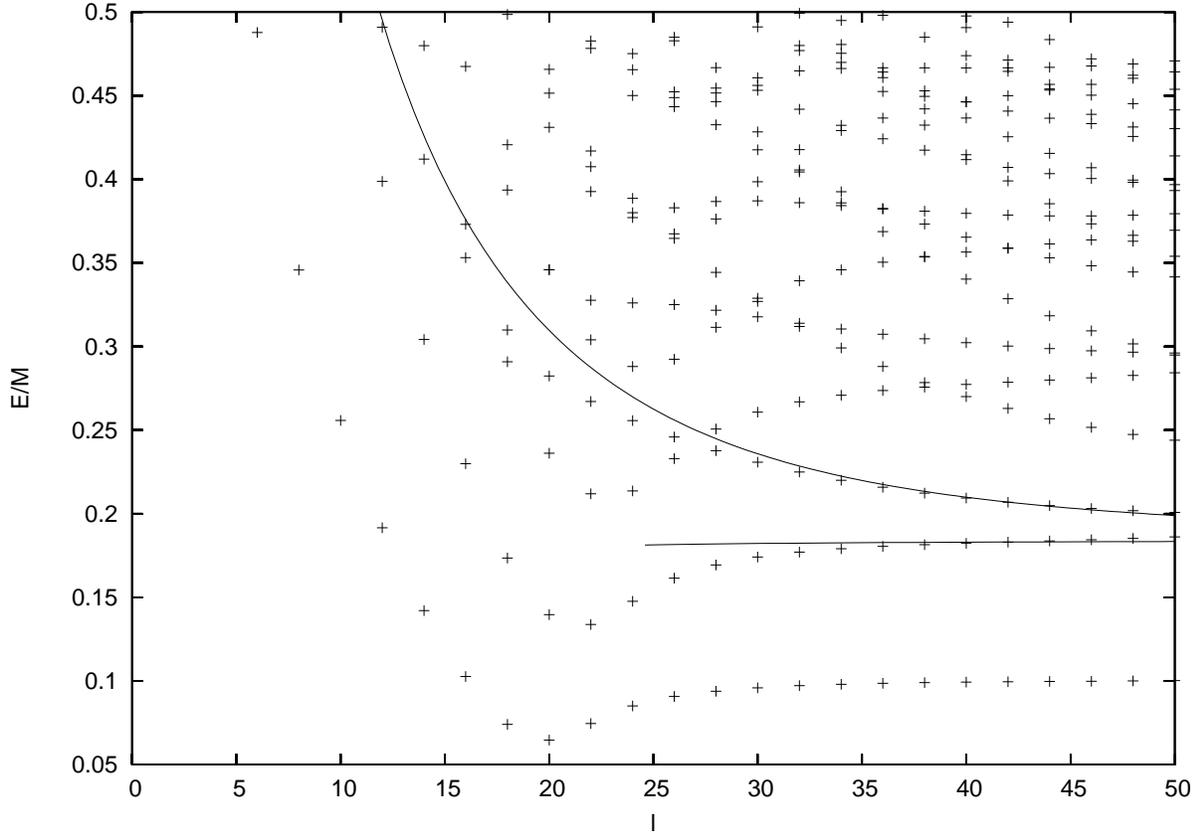}} \par}

\caption{\label{refl_factors2} Boundary excited states at \protect\( \lambda =17\protect \)
and \protect\( \frac{\beta \phi_{0}}{2\pi }=0.485\protect \). The upper line
is the \protect\( I=0\protect \) one-particle \protect\( B_{1}\protect \)
line, including its continuation to imaginary rapidities, while the lower line
is another portion of the imaginary rapidity continuation coming from another
solution of the Bethe-Yang equations. The two lines together fit very well to
the energy level doublet corresponding to the combination of a boundary in its
ground state \protect\( \left| \right\rangle \protect \) and the other in the
excited state \protect\( \left| 0,1\right\rangle \protect \), at least for
sufficiently large values of the volume parameter \protect\( l\protect \). }
\end{figure}

On Figure \ref{refl_factors2}, we illustrate how to obtain
excited boundary states by analytic continuation of one-particle lines.

\subsection{Spectrum of boundary excited states}

We also performed an analysis of boundary excited states for Dirichlet boundary
conditions. As there are two identical boundaries, the states come in doublets
with symmetric/antisymmetric wave functions if the two boundaries are in a different
state, and are singlets if the two boundaries are in the same state. There is
also a selection rule due to a parity introduced by Mattsson and Dorey; namely,
whenever the excited state of the left boundary has an even/odd number of indices,
the right boundary also has even/odd number of indices, respectively. 

For the cases when \( \phi_{0}=\frac{\pi }{\beta } \), the first
excited state is expected to be degenerate with the ground state and
this is indeed what we found within numerical precision. For the other
cases, the energies of the first excited state are summarized in table
\ref{first_excited_dir_table}. This state corresponds to both
boundaries being in the same excited state, so it must be a singlet
and its energy with respect to the ground state (in infinite volume)
is predicted to equal\[ E_{1}-E_{0}=2M\cos \frac{1}{\lambda }\left(
\eta -\frac{\pi }{2}\right) =2M\cos \pi \left( \frac{\lambda
+1}{\lambda }\frac{\beta \phi_{0}}{2\pi }-\frac{1}{2\lambda }\right)
\] We can measure this energy difference using the TCSA data. The
results are illustrated in table \ref{first_excited_dir_table}.

\begin{table}
{\centering \begin{tabular}{|c|c|c|c|}
\hline 
\( \lambda  \)&
\( \frac{\beta \phi_{0}}{2\pi } \)&
\( E_{1}-E_{0} \) (predicted)&
\( E_{1}-E_{0} \) (TCSA)\\
\hline 
\( 31 \)&
\( 0.495 \)&
\( 0.032428 \)&
\( 0.0323(62) \)\\
\hline 
\( 17 \)&
\( 0.485 \)&
\( 0.099750 \)&
\( 0.0997(68) \)\\
\hline 
\( 7 \)&
\( 0.48 \)&
\( 0.14349 \)&
\( 0.143(82) \)\\
\hline 
\( 7 \)&
\( 0.45 \)&
\( 0.35711 \)&
\( 0.357(94) \)\\
\hline 
\( 41/8 \)&
\( 0.44 \)&
\( 0.44675 \)&
\( 0.447(62) \)\\
\hline 
\( 41/8 \)&
\( 0.36 \)&
\( 1.0035 \)&
\( 1.00(64) \)\\
\hline 
\( 17/8 \)&
\( 0.4 \)&
\( 0.89148 \)&
\( 0.89(73)\)\\
\hline 
\end{tabular}\par}

\caption{\label{first_excited_dir_table} Energy of the first boundary excited state
as measured from TCSA}
\end{table}
For higher excited states one can introduce the notion of
level. For a state labeled as \( \left| n_{1},\dots
,n_{k}\right\rangle \) it can be defined as the sum of the integers labels \(
\sum n_{i} \). It turns out that the energies are more or less
hierarchically ordered and increase with the level. We considered
excited states up to and including level \( 2 \) (the first excited
state is at level \( 0 \)) and found excellent agreement with the
predicted spectrum apart from cases when the TCSA spectrum was too
dense to come up with a meaningful identification of the TCSA data
points with individual states. We also fitted them with analytic
continuation of breather lines where this was possible, which also
agreed very well with the TCSA data (see e.g. figure
\ref{refl_factors2}).

\section{\label{sec:conclusions} Conclusions}

In this paper we described an extensive verification of some results
on boundary sine-Gordon theory, comparing numerical TCSA
calculations to predictions concerning the spectrum, scattering
amplitudes, boundary energy and the identification of Lagrangian and
bootstrap parameters of the theory. We found an excellent agreement
and confirmed the general picture that was formed of boundary
sine-Gordon theory in the previous literature.

The main open problems are the calculation of off-shell quantities
(e.g. correlation functions) and exact finite size spectra. While
correlation functions in general present a very hard problem even in
theories without boundaries, in integrable theories significant
progress was made using form factors. One-point functions of bulk
operators have already been computed using form factor expansions in
some simple boundary quantum field theories \cite{bulk_vevs} and one
could hope to extend these results further. In addition, the vacuum
expectation values of boundary operators in sine-Gordon theory are
also known exactly \cite{FZZ}.  It would be interesting to make
further progress in this direction.

Concerning finite size spectra, there is already a version of the
so-called nonlinear integral equation for the vacuum (Casimir) energy
with Dirichlet boundary conditions \cite{leclair}, but it is not yet
clear how to extend it to describe excited states and more general
boundary conditions, which also seems to be a fascinating
problem.

\vspace{0.5cm}

{\par\centering \textbf{Acknowledgments}\par}

We would like to thank P. Dorey, G. Watts and especially Al.B. Zamolodchikov
for very useful discussions. G.T. was supported by a Magyary postdoctoral fellowship
from the Hungarian Ministry of Education. This research was supported in part
by the Hungarian Ministry of Education under FKFP 0178/1999, 0043/2001 and by
the Hungarian National Science Fund (OTKA) T029802/99.


\begin{thebibliography}{1}
\bibitem{Skl} E.K. Sklyanin, \emph{Funct. Anal. Appl.} {\bf 21} (1987)
164. \\
E.K. Sklyanin,  \emph{J. Phys.} {\bf A21} (1988) 2375-2389.
\bibitem{neumann}Z. Bajnok, L. Palla and G. Tak{\'a}cs, 
\emph{Nucl.~Phys.} \textbf{B614} (2001) 405-448,
hep-th/0106069.
\bibitem{bajnok1}Z. Bajnok, L. Palla, G. Tak{\'a}cs and G.Z. T{\'oth}: \emph{The spectrum of boundary
states in sine-Gordon model with integrable boundary conditions,}
preprint ITP-BUDAPEST-571, hep-th/0106070.
\bibitem{ghoshal}S. Ghoshal and A.B. Zamolodchikov, \emph{Int. J. Mod. Phys.} \textbf{A9} (1994)
3841-3886 (Erratum-ibid. \textbf{A9} (1994) 4353), hep-th/9306002.
\bibitem{ghoshal1}S. Ghoshal, \emph{Int. J. Mod. Phys.} \textbf{A9} (1994) 4801-4810, hep-th/9310188.
\bibitem{FK}A. Fring and R. K\"oberle, \emph{Nucl. Phys.} \textbf{B421} (1994)
159, hep-th/9304141.
\bibitem{skorik}S. Skorik and H. Saleur, \emph{J. Phys.} \textbf{A28} (1995) 6605-6622, hep-th/9502011.
\bibitem{bct}P. Dorey, R. Tateo and Gerard Watts, \emph{Phys. Lett.} \textbf{B448} (1999)
249-256, hep-th/9810098.
\bibitem{mattsson}P. Mattsson and P. Dorey, \emph{J. Phys.} \textbf{A33} (2000) 9065-9094, hep-th/0008071.
\bibitem{YZ}V.P. Yurov and A.B. Zamolodchikov, \emph{Int. J. Mod. Phys.} \textbf{A5} (1990)
3221-3246.
\bibitem{dptw}P. Dorey, A. Pocklington, R. Tateo and G. Watts, \emph{Nucl. Phys.} \textbf{B525}
(1998) 641-663, hep-th/9712197.
\bibitem{leclair}A. LeClair, G. Mussardo, H. Saleur and S. Skorik, \emph{Nucl. Phys.} \textbf{B453}
(1995) 581-618, hep-th/9503227.
\bibitem{unpublished}Al.B. Zamolodchikov, unpublished.
\bibitem{FZZ} V. Fateev, A.B. Zamolodchikov and Al.B. Zamolodchikov: \emph{Boundary Liouville
field theory. 1. Boundary state and boundary two point function,} preprint RUNHETC-2000-01,
hep-th/0001012.
\bibitem{Ed}E. Corrigan and A. Taormina \textsl{J. Phys.} \textbf{A33} (2000)
8739. (\texttt{hep-th/0008237}), E. Corrigan \texttt{hep-th/0010094}.
\bibitem{ZZ} A.B. Zamolodchikov and Al.B. Zamolodchikov {\sl
Ann. Phys.} {\bf 120} (1979) 253.
\bibitem{yl_potts_tba}A.B. Zamolodchikov, \emph{Nucl. Phys.} \textbf{B342} (1990) 695-720.
\bibitem{sinhG_tba}Al.B. Zamolodchikov: \emph{On the thermodynamic Bethe Ansatz equation in sinh-Gordon
model}, preprint LPM-00-15, hep-th/0005181.
\bibitem{sinhG_bulk}C. Destri and H. de Vega, \emph{Nucl. Phys.}
\textbf{B358} (1991) 251.
\bibitem{LZ} S. Lukyanov, A.B. Zamolodchikov, \emph{Nucl. Phys.}
\textbf{B493} (1997) 571, hep-th/9611238.
\bibitem{ger}K. Graham , I. Runkel and G. Watts: \emph{Minimal model boundary flows and c=1
CFT}, preprint KCL-MTH-01-01, PAR-LPTHE-01-01, hep-th/0101187.\\
K. Graham, I. Runkel, G. Watts: \emph{Renormalization group flows of boundary
theories,} preprint KCL-MTH-00-54, hep-th/0010082. Talk presented at 4th Annual
European TMR Conference on Integrability Nonperturbative Effects and Symmetry
in Quantum Field Theory, Paris, France, 7-13 Sep 2000.
\bibitem{frt}G. Feverati, F. Ravanini and G. Tak{\'a}cs, \emph{Phys. Lett.} \textbf{B430} (1998)
264-273, hep-th/9803104.\\
G. Feverati, F. Ravanini and G. Tak{\'a}cs, \emph{Nucl. Phys.} \textbf{B540} (1999)
543-586, hep-th/9805117. 
\bibitem{kfold}Z. Bajnok, L. Palla, G. Tak{\'a}cs and F. W{\'a}gner, \emph{Nucl. Phys.} \textbf{B587}
(2000) 585-618, hep-th/0004181.
\bibitem{massgap}Al. B. Zamolodchikov, \emph{Int. J. Mod. Phys.} \textbf{A10} (1995) 1125-1150.
\bibitem{bulk_vevs}R. Konik, A. LeClair and G. Mussardo, \emph{Int. J. Mod. Phys.} \textbf{A11}
(1998) 2765-2782, hep-th/9508099.\\
F. Lesage and H. Saleur, \emph{J. Phys.} \textbf{A30} (1997) L457-463, cond-mat/9608112.\\
 P. Dorey, M. Pillin, R. Tateo and G. Watts, \emph{Nucl. Phys.} \textbf{B594}
(2001) 625-659, hep-th/0007077.

\end{thebibliography}
\end{document}